\documentclass[12pt]{article}
\usepackage{graphicx}
\usepackage{ifpdf}
\ifpdf  \fi
\usepackage{amsfonts,amssymb}
\usepackage{amsthm}
\usepackage{amsmath}
\usepackage[dvipsnames]{xcolor}
\usepackage[ruled,vlined]{algorithm2e} 
\usepackage{newtxtext}

\everymath{\displaystyle} 

\usepackage{subcaption}
\usepackage[colorinlistoftodos]{todonotes}

\usepackage{mathptmx,helvet,courier,makeidx,multicol,footmisc}
\usepackage[numbers]{natbib}
\bibpunct{(}{)}{;}{a}{,}{,}

\usepackage[bookmarks=false, pdfauthor={Irene Martinez}]{hyperref}
    \hypersetup{colorlinks,
      linkcolor=blue,
      citecolor=Emerald,
      urlcolor=blue}

\usepackage[colorinlistoftodos]{todonotes}

\usepackage{nomencl}
\makenomenclature

\newcommand{\commentout}[1]{}

\newcommand{\ba}{\begin{array}}
        \newcommand{\ea}{\end{array}}
\newcommand{\bc}{\begin{center}}
        \newcommand{\ec}{\end{center}}
\newcommand{\bdm}{\begin{displaymath}}
        \newcommand{\edm}{\end{displaymath}}
\newcommand{\bds} {\begin{description}}
        \newcommand{\eds} {\end{description}}
\newcommand{\ben}{\begin{enumerate}}
        \newcommand{\een}{\end{enumerate}}
\newcommand{\beq}{\begin{equation}}
        \newcommand{\eeq}{\end{equation}}
\newcommand{\bfg} {\begin{figure}[h]}
        \newcommand{\efg} {\end{figure}}
\newcommand{\bi} {\begin {itemize}}
        \newcommand{\ei} {\end {itemize}}
\newcommand{\bqn}{\begin{eqnarray}}
        \newcommand{\eqn}{\end{eqnarray}}
\newcommand{\bqs}{\begin{eqnarray*}}
        \newcommand{\eqs}{\end{eqnarray*}}
\newcommand{\bsl} {\begin{slide}[8.8in,6.7in]}
        \newcommand{\esl} {\end{slide}}
\newcommand{\bsq}{\begin{subequations}}
        \newcommand{\esq}{\end{subequations}}       
\newcommand{\bss} {\begin{slide*}[9.3in,6.7in]}
        \newcommand{\ess} {\end{slide*}}
\newcommand{\btb} {\begin {table}}
        \newcommand{\etb} {\end {table}}

\newcommand{\refe}[1] {{(Eq. \ref {#1})}}

\def\pmb#1{\setbox0=\hbox{$#1$}%
   \kern-.025em\copy0\kern-\wd0
   \kern.05em\copy0\kern-\wd0
   \kern-.025em\raise.0433em\box0 }

\newtheorem{theorem}{Theorem}[section]
\newtheorem{definition}[theorem]{Definition}

\newtheorem{lemma}[theorem]{Lemma}
\newtheorem{corollary}[theorem]{Corollary}

\usepackage[draft]{changes}
\definechangesauthor[name={IM},color=red]{IM}

\usepackage[colorinlistoftodos]{todonotes}

\graphicspath{ {figures/} }

\usepackage{multirow}
\usepackage{float}

\usepackage{setspace}   

\begin{document}

\title{Dynamic distance-based pricing scheme for high-occupancy-toll lanes along a freeway corridor}

\author{Irene Mart\'inez and Wen-Long Jin
\thanks{Department of Civil and Environmental Engineering,  University of California Irvine. Corresponding author:  wjin@uci.edu. }
}

\markboth{Mart\'inez and Jin}%
{Shell \MakeLowercase{\textit{et al.}}: A Sample Article Using IEEEtran.cls for IEEE Journals}


\maketitle

\begin{abstract}
   Single-occupancy vehicles (SOVs) are charged to use the highoccupancy-
toll (HOT) lanes, while high-occupancy-vehicles (HOVs) can
drive in them at no cost. The pricing scheme for HOT lanes has been
extensively studied at local bottlenecks or at the network level through
computationally expensive simulations. However, the HOT lane pricing
study on a freeway corridor with multiple origins and destinations as well
as multiple interacting bottlenecks is a challenging problem for which no
analytical results are available.
In this paper, we attempt to fill the gap by proposing to study the  {traffic dynamics in the corridor} based on the relative space paradigm. In this new paradigm, the interaction of multiple bottlenecks and trips can be captured with Vickrey's bathtub model by a simple ordinary differential equation.
   We consider three types of lane choice behavior and analyze their properties.
   Then, we propose a distance-based dynamic pricing scheme based on a linear combination of I-controllers. This closed-loop controller is independent of the model and feeds back the travel time difference between HOT lanes and general-purpose lanes. 
   {Given the mathematical tractability of the system model,} we analytically study the performance of the proposed closed-loop control under constant demand and show the existence and stability of the optimal equilibrium. 
   Finally, we verify the results with numerical simulations considering a typical peak period demand pattern.
   In the future, we are interested in extending this work and testing the performance of the proposed linear combination of I-controllers for other traffic flow models.
\end{abstract}


\section{Introduction}
Hoccupancy vehicle (HOV) lanes were originally designed to encourage ride-sharing and reduce the environmental impacts of congestion  \cite{Shewmake2012}. However, if the HOV lanes are heavily underutilized and general-purpose (GP) lanes are very congested, there is a benefit (from the system perspective) to allowing a certain number of single-occupancy vehicles (SOVs) to travel on the HOV lanes. 
To control how many SOVs use the HOV lanes, a very successful lane management strategy in the United States is high-occupancy-toll (HOT) lane pricing. The operators charge SOVs to use the HOT lanes under this management strategy. 
In addition to achieving better road utilization, the operator is generating a new revenue source that can be used for the maintenance and construction of other HOT lanes or other traffic congestion alleviation initiatives. 
Another fair utilization of the HOT lanes' revenue is to improve local mobility in the community by implementing a better transit system or improving the street network to encourage active mobility.

Over the years, HOT lane pricing has received increasing interest from the implementation field. The first HOT lane in the United States was implemented in 1995 on State Route 91 in California. Soon after, the first HOV-to-HOT conversion project started operation on I-15 near San Diego. Existing HOT lane facilities in the United States as of 2013 are reviewed by Chung \cite{CHUNG2013}, and the rapid increase in HOT lanes implementation in the last decades can be observed in Figure 1 by Wang et al. \cite{Wang2020}. 
The following two paragraphs summarize the existing types of HOT lane pricing and toll collection methods.

The HOT lane pricing can be one of three types: (i) flat rates, which stay constant over time; (ii) scheduled tolls, where tolls vary over time in a predetermined fashion, e.g., depending on the day-of-week and time-of-day; or (iii) dynamic tolls, i.e., a real-time dependent toll, which is ideally adapted to the current traffic conditions  \cite{DePalma2011}. Many dynamic tolling strategies for HOT lanes have been developed in the literature; for a detailed and recent review of dynamic toll pricing, the readers are referred to Lombardi et al. \cite{Lombardi2021}.

Further, there are three types of toll collections methods used on the HOT lanes  \cite{Yang2012}: (i) pass-based toll, where vehicles buy a pass to enter a toll road at any time, e.g., monthly pass; (ii) use-based toll, where vehicles are charged when they use the infrastructure; and (iii) distance-based toll, where the vehicles are charged based on the distance they travel in the toll lane. There are multiple ways of implementing a distance-based toll, some more complex than others. The simplest way is the ``pay per mile'' strategy. Other more complex strategies include: dividing the managed corridor into numerous sections with different ingress/egress points between the HOT and GP lanes and charging a use-based toll on the portions used; or an origin-destination (OD) based toll, where the toll depends on the OD pairs, aiming to limit inequalities across different routes  \cite{Michalaka2013}.
There is a growing interest in distance-based dynamic pricing, which both adapts to real-time traffic and the idea of ``pay-as-you-use’’. {Implementing time-dependent pricing is more equitable than a fixed pricing scheme because it allows for better management of demand and only charging users when needed. Further, distance-based pricing is also a more equitable way of implementing a toll because vehicles that travel short distances generate fewer externalities than vehicles that travel long distances and, thus, should be charged less.}

To start implementing HOT lane pricing, the demand must be high enough so that there is a reason (and social benefit) to charge SOVs. If the demand is so low that the vehicles can be distributed among all lanes without causing congestion, the most equitable toll would be \$0. 
In other words, the main assumption to start implementing HOT lane pricing is related to the demand, and it can be stated as follows:
\begin{itemize}
    \item[A1.] The total demand (HOVs and SOVs) is large enough to generate congestion in the corridor, even if all vehicles were distributed equally among all lanes. The HOVs' demand is assumed to be so low that the HOT lanes are underutilized under no control case. On the other hand, the SOVs demand is assumed to be large enough to lead to a congested state in the GP lanes without control. 
\end{itemize}

In the following, we discuss the operational objectives of HOT lanes for dynamic pricing.
Naturally, the operator of HOT lanes needs to avoid a situation where the HOT lanes become congested since some users are paying for faster speeds. 
Most of the dynamic toll strategies for HOT lanes deployed to date are based on heuristic pricing schemes. For example, the I-15 Express Lane in San Diego uses a pricing mechanism to maintain the level of service C on the express lanes   \cite{Brownstone2003}.
Another example is a recent project to improve the I-405 in Southern California, which includes a new managed lane, where solo drivers will pay the full toll price, and carpools are anticipated to be offered some discounts (or travel for free). This project aims to use dynamic tolls in response to traffic demand. 
The operator's objectives in the I-405 project are to  ``optimize throughput at free-flow speeds'', ``increase average vehicle occupancy'', and ``generate sufficient revenue to sustain the financial viability of the express lanes'', among others  \cite{octa}. The initial toll policy is planned to be a mileage rate varying from \$0.15 to \$0.85 per mile during the non-peak period. The rate will increase in \$0.75 or \$1.00 increments if the hourly flow in the express lane during the peak period is too high  \cite{OCTA_report}. However, flow is not a good indicator of road congestion. For this reason, the flow rate should not be the only reference variable to change the toll price. For example, if operators raise prices when the toll lanes are not busy (low flow), fewer SOVs will pay and switch, worsening congestion on the GP lanes. Thus, better indicators of congestion on the road are speed and density. 

Instead of relying on heuristic strategies, the pricing should be designed to maximize HOT lane utilization. In fact, it is generally assumed ``that as long as there is congestion on the main lanes, the toll can be set so that the HOT lanes are fully utilized, but not congested’’  \cite{Dahlgren2002}. 
This concept of full-utilization of HOT lanes has been adopted and used by several researchers  \cite{Lou2011, Gardner2013}. 
More recently, the HOT lanes pricing operation was studied based on the Point Queue Model (PQM) as a two-fold objective \cite{Jin2020_HOT}. In this paper, we generalize this idea to any traffic model, and we assume:
\begin{itemize}
    \item[A2.] The operational objective for HOT lanes is two-fold. First, the traffic in the HOT lanes should not be congested. Second, the HOT lanes should not be underutilized, i.e., they should be fully utilized. 
\end{itemize}




Considering how the traffic dynamics on managed lanes are modeled before proposing pricing strategies is essential.
Many researchers consider a single bottleneck and use-based tolls  \cite{Lou2011, Gardner2013, Gardner2014, Boyles2015, Wang2020, Jin2020_HOT}. However, these models are simple and cannot be easily extended to study a whole corridor. In fact, very few studies in the literature consider the study of managed lanes for a freeway corridor. We classify these studies based on the geometry considered and the lane choice behavior. 
Regarding road geometry, there are two ways to model HOT lane pricing at corridors. The simplest way is to consider a short stretch of road, where all vehicles have the same trip distance with a single bottleneck. Alternatively, a longer stretch that includes multiple on- and off-ramps can be considered. In this second geometry, vehicles might have different origins and destinations and, consequently, different trip distances. Additionally, there could be multiple bottlenecks that interact with each other.



Most studies considering the first type of geometry, i.e., a short stretch with a single bottleneck, assume that drivers have a single opportunity to choose the HOT lanes. However, some studies have tried to generalize the lane choice behavior by implementing multiple ingress and egress points to the HOT lanes from the GP lanes. Zhu and Ukkusuri \cite{Zhu2015} considered multiple ingress points to the HOT lanes, which allowed them to propose distance-based pricing through a reinforcement learning approach. They relied on the cell transmission model (CTM)  \cite{DaganzoCTM} to capture traffic dynamics. However, the real-time implementation of the algorithm is limited because of the high computational cost. Pandey and Boyles \cite{Pandey2018} further relaxed the assumption of ingress and egress points to the HOT lanes and tested their algorithm through numerical simulations in several networks based on the CTM.

However, since a corridor has multiple interacting bottlenecks and vehicle trips with multiple ODs, the second geometry assumption should be used to study HOT lanes in a corridor. Therefore, 
\begin{itemize}
    \item[A3.]  {We consider corridor stretch with multiple on- and off-ramps, where vehicles might have different origins and destinations and consequently different trip distances. In this geometry setup, multiple bottlenecks are considered in the corridor and can interact with each other.  }
\end{itemize}  
To the best of the authors' knowledge, only three studies have addressed the HOT lane pricing design under this second type of geometry. First, Tan and Gao \cite{Tan2018} proposed a model predictive control and OD-based tolling strategy, i.e., each OD pair might have a different toll at a given time, leading to a mixed-integer linear program. The authors relied on the CTM for the traffic simulations. Proposing an OD-based tolling strategy may lead to unfair pricing since the toll per mile may differ between OD pairs  \cite{Michalaka2013}. Therefore, Tan and Gao  \cite{Tan2018} had to include additional constraints for equity considerations. Second,  Yang et al. \cite{Yang2012} considered a corridor with multiple entrances to the HOT lanes and proposed a distance-based pricing scheme. However, the authors assumed that the congestion on the GP lanes is independent of the number of SOVs that chooses to pay and modeled traffic congestion as a stochastic process. Third, Pandey et al. \cite{Pandey2020} proposed a deep reinforcement learning model, where the tolls are defined based on a feed-forward neural network\footnote{A feed-forward neural network is a network where the information moves only in one direction (from the input nodes to the output nodes) without loops or cycles.}, relying on simulations based on the CTM.


It is clear from the above synopsis that there are no analytical studies on the existence, optimality, and stability of equilibrium states in a closed loop system of HOT lane pricing control on a corridor.
The present paper aims to fill this gap by developing dynamic distance-based pricing for HOT lanes along a freeway corridor, relying on analytical results, which is enabled by the bathtub model  \cite{Vickrey1991, Vickrey2020congestion, ARNOTT2013,jin2020generalized}. 
The bathtub model considers the network to be an undifferentiated unit, which is referred to as a bathtub or reservoir  \cite{Johari2021}. Bathtub models also indirectly assume the same average flow rate, density, and speed along the corridor. These models describe the traffic dynamics in an aggregated way by capturing the initiation, progression, and completion of trips as a simple network queuing model, which has a space dimension relative to the remaining trip distance and is independent of network topology. {In other words, bathtub models allow us to capture supply and demand interactions at complex networks, such as neighborhoods, cities, or freeway corridors. Thus, it can model the traffic congestion patterns when multiple bottlenecks are present with analytical equations. }
Therefore, this new space dimension allows a paradigm shift from the traditional transportation system in absolute space. In this relative space, the OD demand data is not required to model the corridor dynamics as in other traditional models. The aggregate data of trip distance distribution (TDD) and the trip initiation rate are sufficient to define the demand for the bathtub model  \cite{jin2020generalized, Martinez2021}. This approach to studying the HOT lane pricing at the corridor level is conceptually different than previous studies because the CTM is a macroscopic model defined in the traditional absolute space.
Further, the bathtub model is computationally much more efficient than the CTM  \cite{DaganzoCTM} where the entire network is discretized into cells, and the inflow and outflow of each cell need to be calculated.


The relative-space framework allows one to consider multiple lane choice behaviors. For example, the most realistic assumption is when SOV drivers can choose to enter and exit the HOT lanes at multiple points (in time and space) and may travel during a portion of their trip in the HOT lanes and another portion of the trip in the GP lanes. The lane choice modeling based on decision routes (multiple entrances and exit points to the HOT lanes) is an interesting research direction that has not been fully exploited  \cite{Pandey2020}. However, the traditional choice models, such as the multi-nomial logit model, present problems with a strong correlation between alternatives. In the case of drivers choosing the route with multiple entrances and exits to the HOT lane, the problem of multi-nomial logit appears because there is significant overlap among alternatives, i.e., the possible routes share one or a sequence of links. 
Although the assumption of the bathtub model to capture the traffic dynamics enables a simple formulation and solution of complex network dynamic pricing problems, the correlation between route alternatives of the choice model can not be easily solved. For this reason, in this paper, we still consider the simplifying assumption that drivers do not change lane type after they enter the system.
This assumption could be relaxed in future studies, but it is a good starting point. In other words, in this study, we assume that
\begin{itemize}
    \item[A4.] Each driver chooses a type of lane when (s)he enters the corridor and remains in it until (s)he exits the freeway corridor.
\end{itemize}  
Note that distance-based pricing control requires a technology that can measure the distance traveled by each vehicle within the HOT lane. Thus, the HOT lane strategy proposed in this paper can rely either on vehicle-to-infrastructure (V2I) communication or plate recognition technology at all the entrances and exits of the corridor.



The present paper is largely inspired by Jin et al. \cite{Jin2020_HOT}, where the HOT lane pricing problem was also studied through a queuing model. However, the authors focused on the single bottleneck problem and used the PQM, where the space dimension is also irrelevant. In contrast, the bathtub model can capture the interactions among multiple bottlenecks  \cite{Johari2021}.
There are four main differences between Jin et al. \cite{Jin2020_HOT} and this paper.
First, in the PQM, the completion rate of trips under congestion is constant. In contrast, in the bathtub model, the completion rate of trips decreases with increasing densities and is determined endogenously. Therefore, to ensure that the HOT lanes are fully utilized (A2), we need to extend the  concept of \emph{residual capacity} of the HOT lanes \cite{Jin2020_HOT}  to \emph{residual service rate}, which captures the variable capacity.
Second, the lane choice behavior in the PQM is different than that in the bathtub model for two reasons: (i) the bathtub model can be used to study how vehicles change lane types multiple times along the corridor, and (ii) the travel time of a vehicle in the PQM is uniquely determined by its time of arrival to the queue, while in the bathtub model the travel time can be calculated endogenously, and it may increase if traffic congestion worsens.
Third, we propose a distance-based toll instead of a fixed toll (per use). 
Fourth, the closed-loop system obtained from the PQM by Jin et al. \cite{Jin2020_HOT} is mathematically and conceptually different than the one studied in this paper.
Considering all this, the bathtub model offers new possibilities to study HOT lane pricing at freeway corridors with simple, analytically tractable traffic models.




	\begin{table}
	\caption{\small List of abbreviations.} \label{tab:abbreviations}
	\centering
	\small
	\begin{tabular}{ll}\hline
		ATFD &  Approximate Triangular Fundamental Diagram\\
		C &  Critical phase \\
		CTM &  Cell transmission model \\
		GBM &  Generalized Bathtub Model \\
		GP &  General-purpose \\
		HOT &  High-occupancy-toll \\
		HOV &  High occupancy vehicle \\
		NFD &  Network Fundamental Diagram \\
		OD &  Origin-Destination \\
		ODE &  Ordinary differential equation \\
		PDE &  Partial differential equation \\
		PQM &  Point Queue Model \\
		SOC &  Strictly over-critical phase \\
		SOV &  Single Occupancy vehicle \\
		SUC &  Strictly under-critical phase \\
		TDD &  Trip distance distribution \\
		UE &  User Equilibrium \\
		VBM &  Vickrey Bathtub Model \\
		VOT &  Value of time \\
		\hline
	\end{tabular}
	\end{table}

	\begin{table*}
	\caption{\small List of variables.} \label{table:notations}
	\centering
	\small
	\begin{tabular}{lll}\hline
		Variables & Units & Definitions \\\hline
		$c$ &[veh/h]&  Flow-rate per lane under high densities for the ATFD.\\
		$\tilde e_1(t)$ &[veh/h]&  Trip initiation rate of HOVs at $t$.\\
		$\tilde e_2(t)$ &[veh/h] &  Trip initiation rate of SOVs at time $t$.\\
		$\tilde e_{2,1}(t)$ &[veh/h] &  Flux of SOVs vehicles choosing to pay to use HOT lanes at time $t$.\\
		$e_1(t)$ &[veh/h] &  Flux of vehicles entering HOT lanes at time $t$, $e_1(t)=\tilde e_1(t) + \tilde e_{2,1}(t)$. \\
		$e_2(t)$ &[veh/h] &  Flux of SOVs vehicles entering GP lanes at time $t$, $e_2(t)=\tilde e_2(t) - \tilde e_{2,1}(t)$.\\
		$f(\pi)$ &[-] &  Probability density function of value of time (VOT).\\
		$g_i(t)$ &[veh/h] & Completion rate of trips of lane type $i=\{1,2\}$ at $t$.\\
		$l_i$ &[-] &Number of lanes of lane type $i$.\\
		$p(t)$ &[-] & Proportion of SOVs that choose to pay and use the HOT lanes at time $t$.\\
        $q_i(t)$ & [veh/h]  & Flow rate in lane type $i$, a.k.a., internal flow.\\
		$u_f$ &[miles/h]& Free flow speed in the corridor, i.e., parameter of the fundamental diagram.  \\
		$\hat{u}^j$& [\$]& Toll charged to SOV $j$ if it chooses to use the HOT lanes.\\
		$u(t)$&  [\$ / miles] & distance-based toll at time $t$. \\
		$v_i(t)$ &[miles/h] & Average travel speed at time $t$ on lane type $i$.\\
		$w$& [miles/h] & Shock wave speed in congested traffic, i.e., parameter of the fundamental diagram. \\
		$x^j$ &[miles]& Trip distance of vehicle $j$\\\hline 
		$C_0$ &[veh/h]& Road capacity per lane. \\ 
		$C_1$ &[veh/h]& Road capacity of HOT lanes, $C_1 = C_0 l_1$. \\ 
		$C_2$ &[veh/h]& Road capacity of GP lanes, $C_2 = C_0 l_2$.\\ 
		$\tilde D(t)$& [miles]& Average entering trips' distances at $t$. \\
		$E_i(t)$ &[veh]&  Cumulative trip initiation rate at time $t$ in lane type $i$, i.e., $E_i(t) = \int_0^t e_i(\tau) d\tau$.\\
		$G_i(t)$ &[veh]&  Cumulative trip completion rate at time $t$ in lane type $i$, i.e., $G_i(t) = \int_0^t g_i(\tau) d\tau$.\\
		$K_j$ & [-] & Constant parameter of the integral controller. \\ 
		$L_0$ & [miles]& Total linear miles of the corridor\\
		$L_i$ & [miles]& Total lane miles of lane type $i$, $L_i = l_i L_0$. \\
		$V(\rho)$ & [miles/h]& Network speed-density relation. \\\hline
		$\delta_i(t)$& [veh] & Number of active traveling trips at time $t$ on lane type $i$. \\
		$\lambda(t)$ &[veh/miles/lane] & Excess density in the HOT lanes. \\
		$\xi(t)$&  [veh/h/miles] & Residual service rate of HOT lanes. \\
		$\pi$ &[\$/h]& Value of time. \\
		$\rho_1(t)$ & [veh/miles] & Per lane vehicle density in HOT lanes.\\
		$\rho_2(t)$ & [veh/miles] & Per lane vehicle density in GP lanes.\\
		$\rho_c$ & [veh/miles/lane] & Per lane critical density. \\
		$\rho_j$ & [veh/miles/lane] & Per lane jam density. \\ 
		$\hat{\omega}_i^j(t)$ & [h] & Travel time of vehicle $j$ on lane type $i$ at time $t$. \\
		$\hat{\omega}^j(t)$&  [h] & Travel time difference between choosing GP or HOT lanes for vehicle $j$ at time $t$. \\ 
		$\omega(t)$  &[h/miles] & Per-mile travel time difference between choosing GP or HOT at time $t$. \\ \hline
	
	\end{tabular}
	\end{table*}

For the readers' convenience, Table \ref{tab:abbreviations} presents a list of abbreviations. Further, a list of notations is given in Table \ref{table:notations}.

The rest of the paper is structured as follows.
First, the traffic dynamics described by the bathtub model are presented in Section \ref{sec:bathtub}. Second, the lane choice models considered are presented in Section \ref{sec/lanechoice}, and third, the control strategy by the HOT lanes operators is presented in Section \ref{sec/operation}. All these three parts of the HOT lane framework are interconnected: for example, the lane choice of a user depends both on the toll and the traffic state. 
Section \ref{sec:EQHOT}, presents the equilibrium solution and stability analysis of the closed-loop system under the assumption of a constant trip initiation rate. Some numerical cases are presented and discussed in Section \ref{sec:numerical}, and the paper concludes with a short discussion for future steps in Section \ref{sec/discussion}.





\begin{figure*}[t]
\centering
	\includegraphics[width=0.85\textwidth]{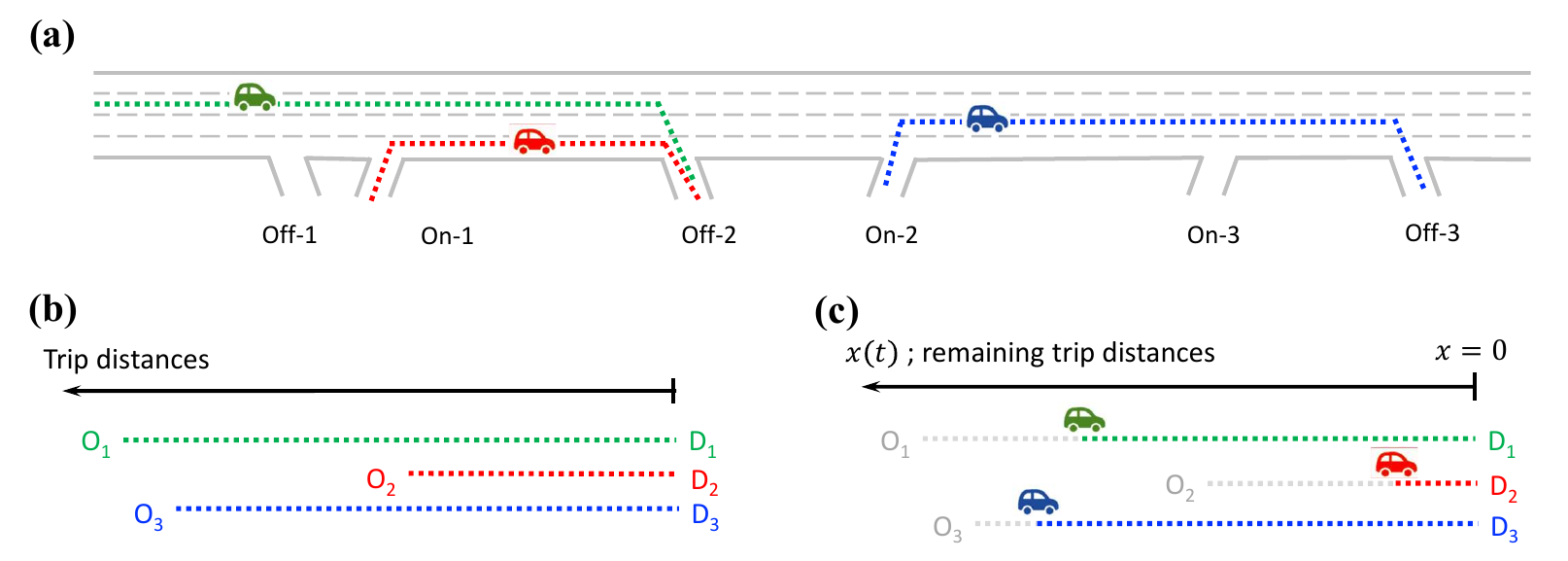} 
	\caption{\small Illustration of a corridor with many on- and off-ramps and subsequent pairs of origin-destination. (a) Corridor geometry with three sample trips; (b) Definition of trip distance; (c) Concept of remaining trip distance. } \label{fig:trips}
\end{figure*}

\section{Vickrey's bathtub model and expected travel time} \label{sec:bathtub}



In this section, we focus on the trip flow dynamics and assume that the result of the lane choice model is given, i.e., the proportion of SOVs that chooses to pay is an input to the bathtub model. 
In the bathtub model, the trips are characterized by their trip distances and trip initiation time. To illustrate the concept of the relative space paradigm, Figure \ref{fig:trips} represents the corridor with a sample of three trips. Figure \ref{fig:trips}(a) presents the trips in the absolute space and Figure \ref{fig:trips}(b) presents the trip distance relative to their destination.  The trip flow dynamics are modeled based on the remaining trip distance to their destination, and trips are completed when their remaining trip distance is $x=0$.

Section \ref{sec/tripflow} presents the details of the bathtub model, including the speed-density relations considered in this paper and other assumptions. Section \ref{sec:expectedtraveltime} presents the approximation considered in this paper on how we estimate the expected travel time. Finally, in Section \ref{sec:relation_variables}, we will present the definition of the state variables in the closed system.

\begin{figure}
\centering
	\includegraphics[width=0.4\textwidth]{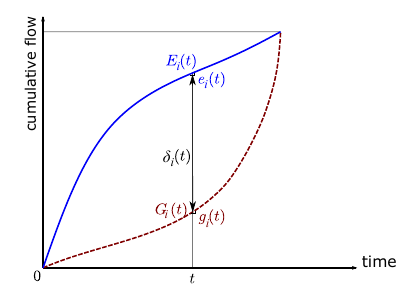} 
	\caption{\small Illustration of the in- and out-flow rates in the bathtub representation of the traffic system with HOT lanes and GP lanes. } \label{fig:bathtubs_ilustration_a}
\end{figure}

\subsection{The bathtub model}\label{sec/tripflow}


Bathtub models  \cite{Vickrey1991, Vickrey2020congestion, ARNOTT2013,jin2020generalized} can be used to study the corridor traffic dynamics with limited computational complexity, by considering a simple queuing system, see Figure \ref{fig:bathtubs_ilustration_a}.
Bathtub models capture the progression of trips by tracking how the remaining trip distance, $x$, is reduced over time depending on the vehicles' speed in the system.  {The bathtub model is assumed as the underlying trip flow model in this study because it represents a good trade-off between model accuracy and mathematical tractability.}

\begin{figure}
\centering
	\includegraphics[width=0.26\textwidth]{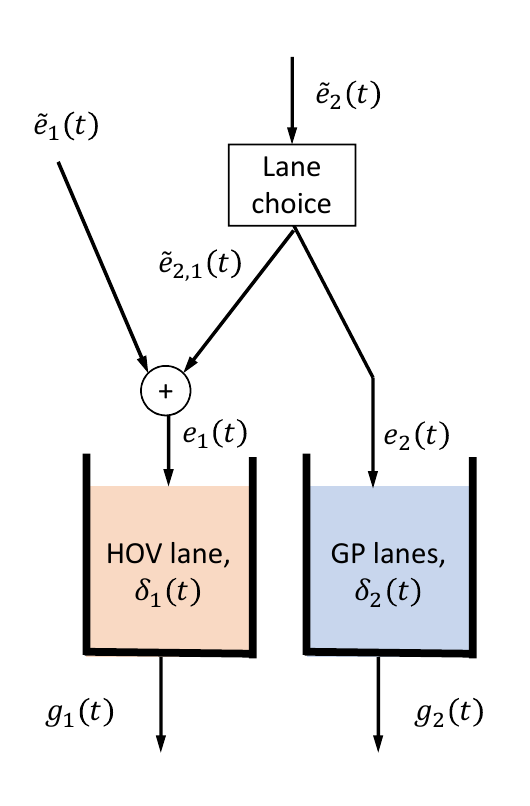} 
	\caption{\small Queuing system for the bathtub model, where the outflow $g_i(t)$ depends on the active number of trips, $\delta_i(t)$, and its remaining trip distance distribution. } \label{fig:bathtubs_ilustration_b}
\end{figure}


In this paper, we assume that the traffic dynamics in each lane type can be modeled with two independent bathtubs, i.e., the traffic dynamics on the HOT lanes and in the GP lanes do not interact. In other words, we assume that the vehicles entering the freeway that chooses to travel in the HOT lanes do not affect the congestion in GP lanes. In reality, the vehicles will need to cross all the GP lanes before entering the HOT lanes, increasing the density temporally on the GP lanes. The impact of these vehicles on congestion is assumed to be negligible, and its actual effect is left for future research.
Further, on-ramps are assumed to have priority. As long as there is space on the road, the entering number of vehicles on the system will be determined by the trip initiation rate, $\tilde e_1(t)$ and $\tilde e_2(t)$ for HOVs and SOVs, respectively. In other words, the demand is exogenous and is not affected by the congestion of the system\footnote{In the future, this study could be extended to include a departure time choice model to determine $\tilde e_1(t)$ and $\tilde e_2(t)$ endogenously.}. Because some SOVs choose to travel in the HOT lanes ($\tilde e_{2,1}(t)$), the inflow to the HOT lanes bathtub, is $e_1 (t) = \tilde e_1(t)+ \tilde e_{2,1}(t)$. Further, the inflow of SOVs to the GP lanes bathtub is $e_2(t) =\tilde e_2(t) - \tilde e_{2,1}(t)$; see Figure \ref{fig:bathtubs_ilustration_b}.

As shown in Figure \ref{fig:bathtubs_ilustration_a}, the rate of change of vehicles in the queue (bathtub) is defined as $ \dot \delta_i(t)$ in lane type $i=\{1,2\}$, where $i=1$ refers to the HOT lanes and $i=2$ refers to the GP lanes. The rate of change of vehicles is the difference between the arrival rate of trips to the corridor (i.e., trip initiation rate) in $e_i(t)$ and the exit rate of vehicles from the corridor (i.e., trip completion rate), $g_i(t)$, for each type of lanes $i$.

As aforementioned, the demand for the bathtub model is not only defined by the trip initiation rate, $e(t)$, but also by the TDD of entering trips. For a freeway corridor, the TDD should be a probability mass function since the entrances and exits of the freeway will determine the possible trip distances. Similar to many other studies in the literature  \cite{Vickrey1991, Vickrey2020congestion, Johari2021}, we assume an exogenous, negative exponential time-independent TDD, with mean trip distance $\tilde D$ for both SOVs and HOVs. {This assumption allows relying on ODEs to model the trip flow dynamics of the bathtub.}
Thereafter, assumption A1 can be particularized for the bathtub model as $\tilde e_1(t) \tilde D  < L_0 C_1$, where $C_1$ is the road capacity of HOT lanes; $\tilde e_2(t) \tilde D > L_0  C_2$, where $C_2$ is the road capacity of the GP lanes; and $(\tilde e_1 (t) + \tilde e_2(t) ) \tilde D > L_0 (C_1 + C_2)$, where $C_i$ represents the road capacity of the lane type $i$.


The bathtub models rely on the assumption that the traveling speed of vehicles $v_i(t)$ can be determined through the Network Fundamental Diagram (NFD), a.k.a. Macroscopic Fundamental Diagram,  \cite{Godfrey1969, Johari2021} as a function of the average per-lane density in the roads, i.e., $v_i(t) = V( \rho_i(t))$, where $\rho_i(t)$ represents the vehicle density in the system, and is related to  the number of active vehicles in a bathtub as
\begin{equation}\label{def/rho}
    \rho_i(t) = \frac{\delta_i (t)}{L_i},
\end{equation}
where $L_i=l_i L_0$ is the total lane miles for a given type of lane.  

 {The completion rate of trips depends on the remaining trip distance of active trips and the speed in the bathtub   \cite{jin2020generalized}. }
Since the TDD is assumed to follow a time-independent, negative exponential distribution, we refer to the bathtub model used as Vickrey's bathtub model (VBM) \cite{Vickrey1991, Vickrey2020congestion}. 
 {Due to the memory-less property of the negative exponential distribution, the} distribution of the remaining trip distance is time-invariant.  {The active trips with no remaining trip distance (i.e., ready to leave the system) is $\delta_i(t) /D $,} where  $D = \tilde D$ is the time-independent, mean remaining trip distance. 
Therefore, the rate of change of active trips in each bathtub can be described with the set of Ordinary Differential Equations (ODEs) as follows
\begin{subequations}\label{eq/dynamics/vbm}
\begin{equation}\label{eq/dynamics/vbm1}
\dot \delta_1(t) = \tilde e_1(t) +  \tilde e_{2,1}(t) - \frac{\delta_1(t)}{D} V\left(\frac{\delta_1(t)}{L_0 l_1}\right) ,
\end{equation}
\begin{equation}
\dot \delta_2(t) = \tilde e_2(t)  - \tilde e_{2,1}(t)  - \frac{\delta_2(t)}{D}V\left(\frac{\delta_2(t)}{L_0 l_2}\right) .
\end{equation}
\end{subequations}
Note that the completion rate of trips  { (a.k.a. exit flow  \cite{Daganzo2007}) $g_i = \frac{\delta_i(t)}{D}V\left(\frac{\delta_i(t)}{L_0 l_i}\right)$ for each bathtub $i$ } differs from the flow rate defined as $q_i(t) = v_i(t) \rho_i(t)$, a.k.a. internal flow   \cite{Daganzo2007}.

\begin{figure*}
    \centering
    \includegraphics[width=0.8\textwidth]{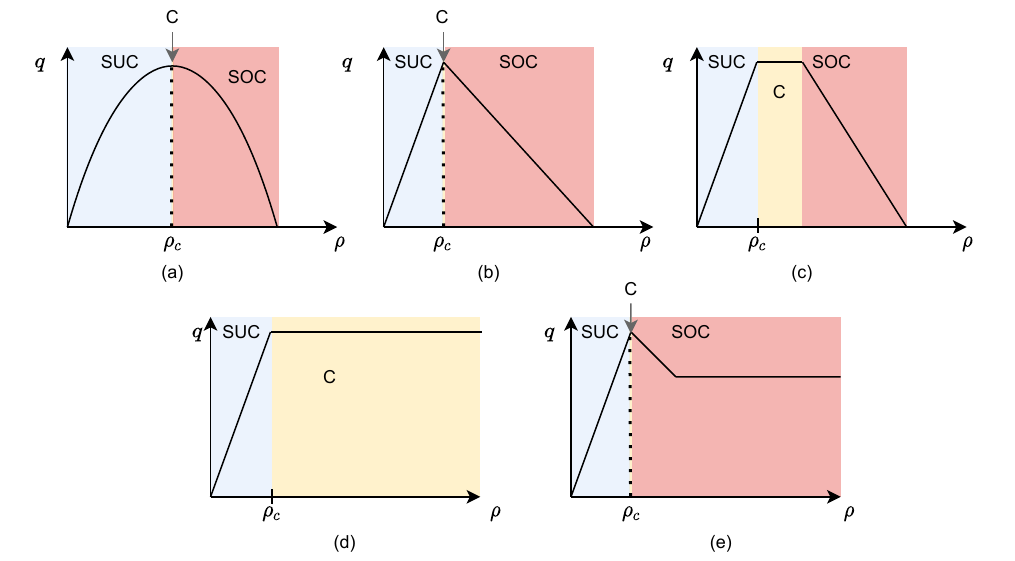}
    \caption{\small Definition of strictly undercritical (SUC), critical (C), and strictly overcritical (SOC) phases in the NFD. For three traditional fundamental diagrams, i.e., (a) Greenshields', (b) Triangular, (c) Trapezoidal, (d) Ramp, and (e) Approximate Triangular.}
    \label{fig:SOCSUC}
\end{figure*}

\begin{definition}\label{def/soc}
Fundamental diagrams have three phases. In the strictly under-critical (SUC), a.k.a. free-flow or under-saturated, the flow is below capacity, and density is lower than the critical density. In the critical phase (C), a.k.a. saturated or congested, the flow is the maximum, i.e., capacity, and the density is critical. In the strictly over-critical phase (SOC), a.k.a. over-saturated or hypercongestion, the flow rate is below capacity, and the density is higher than the critical density. Figure \ref{fig:SOCSUC} presents the three phases described for different fundamental diagrams. 
\end{definition}

Traditionally, the critical density is defined as the density value(s) at which the maximum flow (i.e., capacity) is observed.
For some fundamental diagrams, e.g., the Greenshields' fundamental diagram  (Figure \ref{fig:SOCSUC}(a)) and the triangular fundamental diagram  (Figure \ref{fig:SOCSUC}(b)), the critical density is a unique value. However, for other types of fundamental diagrams, e.g.,  trapezoidal fundamental diagram (Figure \ref{fig:SOCSUC}(c)), the critical density can be a set of connected values. This paper defines the (per lane) critical density, $\rho_c$, as the minimum density at which capacity can be observed. This change in the definition only affects certain NFDs.

\begin{figure*}
\centering
	\includegraphics[width=0.7\textwidth]{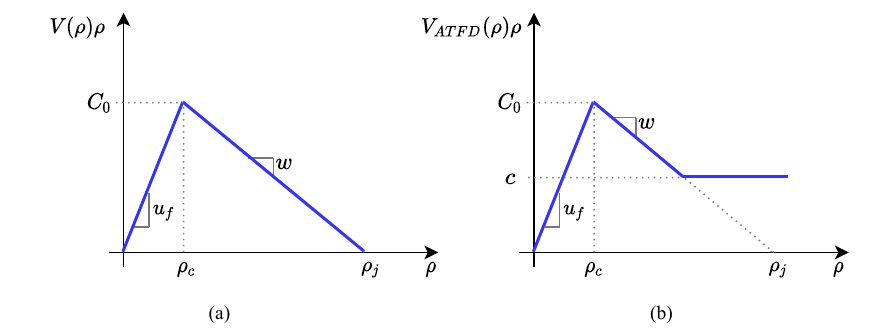} 
	\caption{\small Assumption of (per-lane) fundamental diagram and corresponding main variables. (a) Triangular network fundamental diagram; and (b)  Approximate triangular fundamental diagram (ATFD), where $c$ can range between 0 (equivalent to the triangular fundamental diagram) and $C_0$.} \label{fig:FD} 
\end{figure*}

Traditionally, at  uninterrupted corridors (i.e., no traffic lights) the triangular NFD can be assumed  \cite{Cassidy2011}, i.e.,
\begin{equation}
V(\rho(t)) = \min \bigg\{ u_f , w \left(\frac{\rho_j}{ \rho(t) } - 1 \right) \bigg\}, \label{nfd}
\end{equation}
\noindent
where $u_f$ is the free-flow speed, $w$ is the shock wave speed in traffic congestion, and $\rho_j$ is the per-lane jam density, as depicted in Figure \ref{fig:FD}(a). 
However, the bathtub model with traffic states in  SOC under high inflow leads to hyperconestion, and eventually gridlock  \cite{Fosgerau2015}. \textcolor{black}{To simplify the analysis and ensure system stability under constant demand assumptions in later sections of the paper, we introduce the Approximate Triangular Fundamental Diagram (ATFD). } 
The ATFD is presented in Figure \ref{fig:FD}(b) and the flow-density function can be written as $Q(\rho)=\min \{ u_f \rho , \max \{ w(\rho_j  - \rho ) , c \} \}$; the corresponding  speed-density relation is 
\begin{equation}\label{eq:ATFDspeed}
    V_{ATFD}(\rho) = \min \{ u_f , \max \{\frac{w(\rho_j  - \rho )}{\rho} , \frac{c}{\rho} \} \}.
\end{equation}
\textcolor{black}{The ATFD enables the prevention of gridlock even under high densities, making it a suitable tool for understanding the performance of control strategies.} Note that the AFTD in SOC is, in general, not monotone decreasing. However, taking the limit of $c \longrightarrow 0$, the ATFD leads to the triangular network fundamental diagram, Eq. \ref{nfd}. Further, the particular case of $c=C_0$ is referred to as \emph{ramp} network fundamental diagram (by its resemblance to the ramp function)\footnote{Note that the flow rate is constant both for the ramp NFD at SOC states or the ATFD at very large densities. In those states, the system dynamics will be similar to considering a PQM.}. In all cases, the critical density is defined as
\begin{equation*}
    \rho_c = \frac{w \rho_j}{u_f + w}.
\end{equation*}


\subsection{Expected travel time as an instantaneous approximation}\label{sec:expectedtraveltime}

We define the average travel time of users over a give period of time on a given bathtub $i$, as  $\bar{\hat{\omega}}_i(t)$,  for the HOT lanes ($i=1$) and the GP lanes ($i=2$), respectively. One can obtain this average travel from the cumulative arrival and departure curves to the queuing system presented in Figure \ref{fig:bathtubs_ilustration_a}. The area between the cumulative curves represents the total travel time $ \int_0^T  E_i(\tau) - G_i(\tau) d \tau$,
where $E_i(t)=\int_0^t e_i(\tau) d \tau$, and $G_i(t)=\int_0^t g_i(\tau) d \tau$. 
Then, the average travel time for users in each type of lane is
\begin{equation*}
    \bar{\hat{\omega}}_i = \frac{1}{G_i(T)} \int_0^T E_i(\tau) - G_i(\tau) d \tau.
\end{equation*}

An individual vehicle $j$ entering the system at time $t_j$ cannot know with certainty its expected travel time for each possible lane chosen. In fact, the speed it will experience depends on the active traveling trips in the bathtub after $t_j$. Note that SOVs arriving at  $t> t_j$ will also need to make a lane choice, which will impact the active traveling trips in each bathtub. In other words, the travel time depends on how the system evolves in the future because the completion rate of trips, $g_i(t)$, depends on the vehicles in the system $\delta_i(t)$. 
For these reasons, the travel time information for each lane given to the drivers (either through V2I or through variable message signs) can only be an estimate of the actual travel time the users will experience.
The travel time for individual $j$ choosing lane type $i$, $\hat{\omega}_i^j$ is approximated at a given time $t$ as follows
\begin{equation}\label{omega/approx}
\hat{\omega}_i^j \sim \hat{\omega}_i^j(t) = \frac{x^j}{v_i(t)},
\end{equation} 
\noindent
where $x^j$ is its trip distance, and $v_i(t)$ is the average speed in bathtub $i$ at time $t$. Notice that a vehicle can approximate its travel time at any time while (s)he is in the system, not only at the start time. 

For simplicity, in this paper, we use this instantaneous approximated travel time when users start their trip, $t_j$. However, note that the operator could estimate the expected average speed using other methods instead of the instantaneous speed. For example, the operator could rely on day-to-day dynamics. Although considering these day-to-day dynamics is out of the scope of this paper, it would be an interesting future research direction since it might allow proposing more equitable pricing schemes. For example, to define the pricing scheme, one could consider the marginal congestion increase (e.g., externalities) caused by a driver based on expected travel time and the effect of a driver's decision on the future speeds in the bathtub \cite{Vickrey1963Pricing,VICKREY2019}.

We define the travel time difference between lane types for vehicle $j$ as $ \hat{\omega}^j(t)= \hat{\omega}_2^j(t) - \hat{\omega}_1^j(t)$. 
Then, we define the travel time per unit distance as $\omega_i (t)= \frac{\hat{\omega_i}^j(t)}{x^j}$, and from  Eq. \ref{omega/approx} we have the travel time difference per unit distance as
\begin{equation}\label{omega/def}
    \omega(t)=\left( \frac{1}{v_2(t)} - \frac{1}{v_1(t)} \right),
\end{equation}
which is vehicle-independent, i.e., only depends on the instantaneous speeds of the bathtubs.

\subsection{Definition of state variables}\label{sec:relation_variables}

We define two state variables for the HOT lane pricing scheme. The \textit{excess density} in the HOT lanes as $\lambda(t) = \rho_1(t) - \rho_c$, and from Eq. \ref{def/rho} we have
\begin{equation}\label{def/lan}
    \lambda(t) = \frac{\delta_1(t)}{L_1} - \rho_{c}.
\end{equation}

The  \textit{residual service rate} in the HOT lanes is defined as $\xi(t) = g_1(t) -  e_1(t)$. 
This residual service rate is an extension of  the concept of residual capacity from the PQM by Wang et al. \cite{Wang2020}, $ C_1 - e_1(t)$. 
The extension is needed because the bathtub model does not have a constant trip processing capacity. \textcolor{black}{Physically, the residual service rate is the capability of the HOT lanes to process more trips than the entering trips $e_1(t)$ at a given time.}
From Eq. \ref{eq/dynamics/vbm1} and Eq. \ref{def/lan} we have
\begin{equation}\label{def/xi}
    \xi(t) =(\lambda(t) + \rho_c )  \frac{L_1}{D} V(\lambda(t) + \rho_c) -  \tilde e_1(t) - \tilde e_{2,1}(t),
\end{equation}
which depends on $\lambda(t)$, the mean remaining trip distance, and the assumed NFD.  From Eq. \ref{eq/dynamics/vbm}, Eq. \ref{def/lan}  and Eq. \ref{def/xi} we have 
\begin{equation}\label{def:xi_lan}
    \xi(t) = - L_1 \dot \lambda(t).
\end{equation}


\section{Lane choice models}\label{sec/lanechoice}


In this section, we consider several lane choice models to determine the proportion of SOVs that choose to pay, 
\begin{equation}\label{def/p}
    p(t) = \frac{\tilde e_{2,1}(t)}{\tilde e_2(t)},
\end{equation}
where $\tilde e_{2,1}(t)$ is the number of SOVs that choose to pay, and $\tilde e_2(t)$ is the number of entering SOVs at a given time. The lane choice decision of the SOVs drivers will depend not only on the distance-based toll, $u(t)$, but also on individuals' value of time (VOT) and trip distance, and the speeds on each bathtub.

\subsection{Relation of $p(t)$ to state variables}\label{sec/relationstate}

In the following we discuss the relation between $\lambda(t)$ and $\xi(t)$ with  the proportion of SOVs that choose to pay, $p(t)$.

From  Eq. \ref{def/lan} , Eq. \ref{def/xi}, and Eq. \ref{def/p}  we have
\begin{equation}\label{eq/pxi}
    p(t) = \frac{ (\lambda(t) + \rho_c ) \frac{L_1}{D} V(\lambda(t) + \rho_c ) - \tilde  e_1(t) - \xi(t)}{\tilde e_2(t)}.
\end{equation}

\begin{theorem}\label{theorem/partialp}
For constant density in the HOT lanes and constant trip initiation rates of SOVs and HOVs, an increase in residual service rate is associated with a decrease in paying SOVs, i.e.,  $ \frac{\partial p}{\partial \xi} < 0$, from Eq. \ref{eq/pxi}.

On the other hand, for a constant residual service rate, $ \frac{\partial p}{\partial \lambda}$ depends on the shape of the considered NFD and $\lambda(t)$. Following Definition \ref{def/soc}, 
\begin{itemize}
    \item For SUC we have $ \frac{\partial p}{\partial \lambda} >0$.
    \item For C we have  $ \frac{\partial p}{\partial \lambda} =0$.\footnote{Note that in some piece-wise defined fundamental diagrams (e.g., the triangular and ATFD) $ \frac{\partial p}{\partial \lambda}$ at C is a set of values, where 0 belongs to the set.} 
    \item For SOC we have $ \frac{\partial p}{\partial \lambda} \leq 0$.
\end{itemize}
\end{theorem}

The proof of Theorem \ref{theorem/partialp} is straightforward and is omitted here.

\begin{corollary}\label{corol}
For fundamental diagrams with monotone decreasing flow with density at SOC,  we have $ \frac{\partial p}{\partial \lambda} < 0$. However, for SOC with not a monotone decreasing flow with density, $ \frac{\partial p}{\partial \lambda}$ may be piece-wise defined.
\end{corollary}

An example of a non-monotone decreasing flow with the density of Corollary \ref{corol} is the case of ATFD  Eq. \ref{eq:ATFDspeed}, where the SOC for large enough densities corresponds to constant flow. For the ATFD, we have $ \frac{\partial p}{\partial \lambda} \leq 0$ for $\lambda(t) <(\rho_j - \rho_c)l_1 - \frac{c_1}{w}$, but for $\lambda(t) \geq (\rho_j - \rho_c)l_1 - \frac{c_1}{w}$ we have $ \frac{\partial p}{\partial \lambda} =0$.

Let us provide a physical interpretation of why the partial derivative $\frac{\partial p}{\partial \lambda} $ changes the sign for different regimes of traffic. In SUC phases, higher $\lambda(t)$ leads to higher flow, i.e., higher $g(t)$. Since it is a partial derivative, $\xi(t)$ is assumed to be constant. Thus, to maintain the same residual service rate, the inflow must be larger. Further, if $\tilde e_i(t)$ are constant, a higher inflow can only be related to a higher proportion of SOVs choosing to pay. On the other hand, in a SOC phase, an increase in $\lambda(t)$ leads to a reduction in flow and consequently lower $g(t)$. Thus, to maintain $\xi(t)$ constant, the inflow should be reduced, i.e., $p(t)$ decreases.

\subsection{Properties of lane choice models}\label{sec:lanechoiceproperties}

The VOT can be very different across vehicle types and trip purposes. For example, a commuter VOT is estimated to be about 50\% of their wage, while trucks have a VOT of \$40 per hour  \cite{Casady2020}. In the literature, many different distributions have been considered for the VOT, e.g., uniform  \cite{VandenBerg2011}, lognormal  \cite{Bhat2000}, a simplified Burr distribution  \cite{Gardner2013} and triangular or Johnson's distribution  \cite{HESS2005}. 

This paper will assume a generic time-independent VOT distribution, which we denote as $f(\pi)$.
We define $\pi^j$ as the user's $j$ VOT  and  $\hat{u}^j(t)$ the toll for SOV $j$ if (s)he decides to pay when (s)he starts the trip at time $t$. From Eq. \ref{omega/approx} the disutility functions for SOV $j$ can be defined as follows.  The total cost for SOV $j$ to use the HOT lanes is
\begin{subequations}\label{eq/utility}
\begin{equation}\label{cost/HOT}
   U^j_{1} = \hat{u}^j(t)  + \frac{x^j}{v_1(t)}\pi^j ;
\end{equation}
\noindent
where $x^j$ is the trip distance of user $j$.
Similarly,
\begin{equation}\label{cost/GP}
    U^j_{2} = \frac{x^j}{v_2(t)}\pi^j  ,
\end{equation}
\end{subequations}
\noindent is the cost of the same vehicle $j$ staying in the GP lanes. 


\begin{definition}\label{definition}
Assuming that the drivers' VOT follows a distribution of $f(\pi)$, the lane choice model is well defined if vehicle $i$ chooses the HOT lanes when $\pi^i>\pi^j$ and vehicle $j$ chooses the HOT lanes. Analogously, if $\pi^i<\pi^j$ and vehicle $j$ chooses the GP lanes, then vehicle $i$ has to choose the GP lanes. 
\end{definition}

We refer to this Definition \ref{definition} as the consistency principle for lane choice behaviors. Many choice models can be considered to model the driver's behavior. 

\begin{lemma}\label{lemma/mileage}
A sufficient condition to have a well-defined lane choice model is that the toll per unit distance is independent of the vehicle characteristics, i.e., the same distance-based toll is applied to trips with different trip distances.
\end{lemma}
\begin{proof}
If vehicle $j$ chooses the HOT lanes, $\hat{u}^j(t) + \hat{\omega}_1^j(t)\pi^j < \hat{\omega}_2^j(t)\pi^j  $. From Eq. \ref{omega/def}  we have $\frac{\hat{u}^j(t)}{x^j} < \omega(t) \pi^j$.  Then, if $\pi^i>\pi^j$ we have $\frac{\hat{u}^j(t)}{x^j} < \omega(t) \pi^j < \omega(t) \pi^i$. Thus, if $u(t)=\frac{\hat{u}^j(t)}{x^j}$ is independent of vehicles, $u(t) =\frac{\hat{u}^j(t)}{x^j} < \omega(t) \pi^i$ , and vehicle $i$ will also choose to pay.
Similarly, if vehicle $j$ chooses the GP lanes and $\pi^i< \pi^j$, we have $u(t) > \omega(t) \pi^j > \omega(t) \pi^i$, which leads to $u(t)  > \omega(t) \pi^i$, and $i$ chooses the GP lanes.
\end{proof}

Given Lemma \ref{lemma/mileage}, we define $u(t)$ as the distance-based toll, analogously to the travel time per unit distance,
\begin{equation*}\label{def/mileagetoll}
    u(t) = \frac{\hat{u}^j(t)}{x^j} ,
\end{equation*}
which is trip distance independent. Therefore,  Eq. \ref{eq/utility} differs slightly from the standard formulation of utility function, $U^j_i= \sum_k \theta_{ki}^j x^j_{ki}$, for the alternative $i$ of user $j$. Here, constant linear coefficients are defined as the trip distance $\theta_1=x^j$ and the product of VOT with trip distance $\theta_2=x^j \pi^j$; while the observed variables are independent of the individuals $X_{1,HOT}=u(t)$, $X_{2,HOT}=\frac{1}{v_1(t)}$, $X_{1,GP}=0$ and $X_{2,GP}=\frac{1}{v_2(t)}$. In other words, we consider that the observations are the same for all individuals, but the parameters are randomly distributed among individuals.

Similar to Jin et al. \cite{Jin2020_HOT} , the choice model at an aggregated level can be defined as a function of $u(t)$, $\omega(t)$, and the vehicle's VOTs distribution $f(\pi)$, i.e.,
\begin{equation}\label{def/sig}
    p(t)  =\sigma(\omega(t),f(\pi),u(t)),
\end{equation}
\noindent
where the first variable describes the state of the system and is vehicle independent,  $f(\pi)$ is a characteristic of the demand and is assumed to be independent of the state of the system, and $u(t)$ is the distance-based toll defined by the operator. 
 
Besides Definition \ref{definition}, we also adopt other reasonable behavioral principles for the lane choice model  \cite{Jin2020_HOT}:
\begin{itemize}
    \item [1.] The proportion of SOVs that choose to pay, $p(t)$ decreases with $u(t)$, i.e.
    \begin{subequations}\label{behaviors}
    \begin{equation}\label{behavior1}
        \frac{\partial p}{\partial u} < 0
    \end{equation}
    \item [2.] The proportion of SOVs that choose to pay, $p(t)$, increases with the time difference ${\omega}(t)$. That is, the larger the waiting time difference is, the more SOVs will choose the HOT lanes.
    \begin{equation}\label{behavior2}
        \frac{\partial p}{\partial  \omega} > 0
    \end{equation}    
    \end{subequations}
    \item [3.] For ${\omega}(t)=\infty$, $p(t)=1$; and for $u(t)=\infty$, $p(t)=0$.
\end{itemize}

From  Eq. \ref{eq/pxi} and Eq. \ref{def/sig} we can establish a relation between the distance-based toll and the state variables
\begin{equation*}
    u(t) = \sigma^{-1} (\omega(t),  f(\pi), \frac{ (\lambda(t) + \rho_c )  \frac{L_1}{D} V(\lambda(t) + \rho_c ) - \tilde  e_1(t) - \xi(t)}{\tilde e_2(t)}),
\end{equation*}
\noindent which is an equivalent form of the lane choice model. We assume that the operator does not know the lane choice behavior of SOVs and cannot use the above expression \cite{Jin2020_HOT}.    
A general form of the above expression can defined as $u(t) = M(\omega(t),  f(\pi), \xi(t), \lambda(t))$.

A good understanding of the relation between the $u(t)$ and the state variables is needed to propose a well-behaved pricing scheme.
Therefore, we will consider three different lane choice models where each individual vehicle tries to minimize its cost: (i) drivers are assumed to have a perfect knowledge of the costs, leading to vehicle-based user equilibrium (UE);  (ii) assuming that the individuals do not have a perfect knowledge of the cost, i.e., mixed logit model that leads to a stochastic user equilibrium, and (iii) a generalized lane choice model.

\subsection{Deterministic: Individuals' UE}\label{sec:UE}

Here we assume that a driver chooses the lane type to minimize her/his cost, which is perfectly defined by the disutility function Eq. \ref{eq/utility}, and that (s)he has a perfect knowledge of the different costs. Therefore, a SOV chooses to drive in the HOT lanes if \ref{cost/HOT} is lower than the cost of staying in GP lanes \ref{cost/GP}, or, equivalently if 

\begin{subequations}\label{conditions}
\begin{equation} \label{cond:HOT}
    u(t)  \leq \pi^j  \omega(t),
\end{equation}
\noindent
based on Eq. \ref{omega/def}. Note that $\pi$ is a random variable. Recall that $\omega(t)$ is an approximation and indirectly assumes that the traffic conditions, e.g., the speed of the system, will not change during vehicle $j$'s trip. 
In contrast, if the cost of vehicle $j$ taking the HOT lanes is larger than staying in the GP, that is $    \hat{u}^j(t) + \hat{\omega}_1^j(t)\pi^j > \hat{\omega}_2^j(t)\pi^j $, vehicle $j$ will choose to stay in GP lanes. From Eq. \ref{omega/def} this is equivalent to
\begin{equation}\label{cond:GP}
     u(t)  > \pi^j  \omega(t).
\end{equation}
\end{subequations}
Therefore, we have
\begin{equation*}
p(t)=Pr\{\pi^j | \pi^j \geq \Pi(t)\equiv \frac{u(t)}{\omega(t)}  \}    .
\end{equation*}
\noindent

For this choice model to be consistent for all vehicles, $\Pi(t)$ should be independent of vehicles, which is satisfied if $u(t)$ is vehicle independent, as discussed in Lemma \ref{lemma/mileage}. 
Then, for a set of vehicles entering the system simultaneously, the ones with larger VOT, $\pi^j$, will be more likely to choose to pay for the HOT lanes.
This intuition is reasonable for real-life behaviors. Thus, the proposed lane choice model is an adequate behavioral assumption. 

To determine the proportion of vehicles that will choose to use HOT lanes, we need to integrate the distribution $f(\pi)$ over the VOTs that satisfy Eq. \ref{cond:HOT}, which is
\begin{equation*}
    p(t) = \int_{\frac{u(t)}{\omega(t)}}^{\infty} f(\pi) d \pi.
\end{equation*}
\noindent
By definition of the cumulative density function, we have
\begin{equation}\label{def/pUE}
    p(t) = 1 - F\left(\frac{u(t)}{\omega(t)}\right).
\end{equation}
\noindent Notice that this mathematical expression is equivalent to the bottleneck case scenario first presented by Gardner et al. \cite{Gardner2013} and first related to the UE principle by Jin et al. \cite{Jin2020_HOT}. The expressions are equivalent because both $u(t)$ and $\omega(t)$ here are defined per unit distance, and thus their ratio is the same as the total price over the travel time difference. 
It is clear that Eq. \ref{def/pUE} satisfies the reasonable behavioral principles described by Eq. \ref{behaviors}.


Further, Eq. \ref{def/pUE} can be rewritten to capture the distance-based price $u(t)$ as
$ u(t) = \omega(t) z(p)$,
where $z(p)$ is defined as the 100(1-$p$)-th percentile defined by $p= 1 - F(z(p))$. 
From Eq. \ref{eq/pxi} we have
\begin{equation*}
    u(t) = \omega(t) z(\frac{ (\lambda(t) + \rho_c ) \frac{L_1}{D}V(\lambda(t) + \rho_c ) - \tilde  e_1(t) - \xi(t)}{\tilde e_2(t)}),
\end{equation*}
which corresponds to the particular function $M(\omega(t),  f(\pi), \xi(t), \lambda(t))$ for the UE lane choice model. For different assumptions of $f(\pi)$, the lane choice model will have different mathematical expressions, which can be generalized as
\begin{equation}\label{gen/UE}
     u(t) = \omega(t) A( \xi(t) , \lambda(t)).
\end{equation}

\begin{lemma}\label{lemma/A}
$A(\xi(t), \lambda(t))$ is non-negative.
\end{lemma}
\begin{proof}
 Notice that $z(p)\geq0$ by definition, and thus we have
\begin{equation}\label{eq/uw}
    \frac{\partial u}{\partial \omega } \geq 0.
\end{equation}
\end{proof}

From Theorem \ref{theorem/partialp} we have
\begin{subequations}\label{condition1}
\begin{equation}\label{eq/Axi}
     \frac{\partial A}{\partial \xi } > 0,
\end{equation}
because  $\frac{\partial z}{\partial p } < 0 $ holds by definition.
Eq. \ref{eq/Axi} indicates that the distance-based toll and $\xi(t)$ are positively correlated. In other words, to increase the proportion of SOVs paying when the residual service rate is large, the operator should reduce  $u(t)$. This relation is consistent with Eq. \ref{behavior1}, the first reasonable behavioral principles. Further, from Theorem \ref{theorem/partialp} we have
\begin{equation}
     \frac{\partial A}{\partial \lambda } < 0 \text{ for SUC, }  \frac{\partial A}{\partial \lambda } \geq 0 \text{ for SOC, and } \frac{\partial A}{\partial \lambda } = 0 \text{ for C}.
\end{equation}
Note that $\frac{\partial A}{\partial \lambda }$ is not necessarily the same sign as $\frac{\partial u}{\partial \lambda }$, because
\begin{equation}
    \frac{\partial u}{\partial \lambda } = \frac{\partial \omega(t)}{\partial \lambda } A( \xi(t) , \lambda(t)) + \omega(t) \frac{\partial A ( \xi(t) , \lambda(t) )}{\partial \lambda },
\end{equation}
where the first term is always negative, and the second depends on the traffic phase (SOC, SUC, or C). Thus, to increase the excess density on SUC, the operator should reduce $u(t)$. In contrast, to reduce the excess density on SOC, the operator might increase or decrease $u(t)$, depending on the other variables.
\end{subequations}


\subsection{Probabilistic: Logit model}\label{sec:logit}


In this case, we assume that SOVs choose to pay and use the HOT lanes based on the logit model.
The logit model is a discrete choice model based on random utility maximization \textcolor{black}{and assumes that the perceived (dis)utility of users is not known by the modeler. In other words, } Eq. \ref{eq/utility} is \textcolor{black}{assumed to be an inaccurate} description of the individuals' disutility\textcolor{black}{, because} there are some individuals' characteristics that affect the lane choice that the modeler cannot capture, \textcolor{black}{for example,} the travel time reliability. \textcolor{black}{To account for this, the logit model introduces an error term (random component) to the disutility, which represents the unobserved portion of the disutility.} The binary logit model assumes the random components of the cost of the different alternatives, $\epsilon_i$, are independent and identically distributed (i.i.d.) with a Gumbel distribution; and that the error variance-covariance structure of the alternatives is identical across individuals.

Moreover, we assume \textcolor{black}{that the discrepancy between the modeled disutility and the actual disutility is proportional} to the trip distance, i.e., the disutility is $U_i^j + x^j \epsilon_i$. 
We then define a per unit distance disutility that includes the unknown error as

\begin{equation*}
    W_1^j = u(t) + \frac{\pi^j}{v_1(t)}  + \epsilon_1, \text{ and}
\end{equation*}
\begin{equation*}
    W_2^j = \frac{\pi^j}{v_2(t)}  + \epsilon_2,
\end{equation*}

for the HOT lanes and GP lanes, respectively. Therefore, the probability that a vehicle $j$ chooses to use the HOT lanes is
\begin{equation*}
   P\{j \text{ chooses HOT lanes} \} = \frac{1}{1 + e^{\alpha_* \left[ u(t) - \pi^j \omega (t) \right]  } },
\end{equation*}
\noindent
where $\alpha_*$ is the scale parameter. To determine the proportion of vehicles that choose to pay under these assumptions, we need to rely on the mixed logit model  \cite{McFadden2000} because the VOT $\pi^j$ \textcolor{black}{may be} different across the population. However, we can consider the simplifying assumption that all users have the same VOT, i.e., $\pi^j=\pi^*$ for all $j$. Then, the proportion of paying SOVs is
\begin{equation}\label{eq/pML}
    p(t) =\frac{1}{1 + e^{\alpha_* \left[ u(t) - \pi^*  \omega (t) \right]  } } .
\end{equation}

\noindent which 
can be written as
\begin{equation}\label{eq/u/logit}
    u(t) = \omega(t) \pi^* + \frac{1}{\alpha_* } \ln \left( \frac{1}{p(t)} - 1 \right).
\end{equation}
From Eq. \ref{eq/pxi}, we have the following equivalent lane choice model based on the VBM trip flow dynamics
\begin{equation} \label{eq/ulogit}
\begin{split}
    u(t) & = \\
    &\frac{1}{\alpha_* } \ln \left( \frac{\tilde e_1(t) + \tilde e_2(t) +\xi(t) - (\lambda(t) + \rho_c ) \frac{L_1}{D} V(\lambda(t) + \rho_c )}{(\lambda(t) + \rho_c ) \frac{L_1}{D} V(\lambda(t) + \rho_c ) - \tilde e_1(t) - \xi(t)} \right)\\
    & + \omega(t) \pi^*,
\end{split}
\end{equation}
\noindent which can be generalized as 
\begin{equation}\label{gen/logit}
    u(t) = \omega(t) \pi^* +B(\xi(t), \lambda(t)).
\end{equation}
\noindent From Eq. \ref{gen/logit}, $\frac{\partial u}{\partial \omega}$ is non-negative as in the UE model Eq. \ref{eq/uw}.
\begin{subequations}\label{eq/relations4}
From Eq. \ref{eq/ulogit} we have
\begin{equation}\label{eq/Dxi}
    \frac{\partial B}{\partial \xi} >0.
\end{equation}
Since $\omega(t) \pi^* $ is not dependent on $\xi(t)$, if the operator aims to reduce the residual service rate $\xi(t)$, (s)he should reduce $u(t)$. 
This is consistent with the first of the reasonable behavioral principles defined in previous sections (Eq. \ref{behavior1}) and the results of the UE (Eq. \ref{condition1}).  Further, from   Eq. \ref{eq/ulogit} and Theorem \ref{theorem/partialp} we have 
\begin{equation}
    \frac{\partial B}{\partial \lambda} <0 \text{ for SUC, } \frac{\partial B}{\partial \lambda} \geq 0 \text{ for SOC, and } \frac{\partial B}{\partial \lambda} = 0 \text{ for C.}
\end{equation}
\end{subequations}
Note that the relation of $B(\xi(t), \lambda(t))$ in the logit model has the same relation as $A(\xi(t), \lambda(t))$ in the UE model to $\lambda(t)$, see Eq. \ref{condition1}. Again, these relations do not show how  the toll $u(t)$ and $\lambda(t)$ are related, except for the SUC phase, where  $\frac{\partial u}{\partial \lambda} <0$.


\subsection{Generalized lane choice model}


Considering the two very particular cases above,  (i) the UE lane choice model with distributed VOT, $f(\pi)$, and (ii) the logit choice model with deterministic  VOT, we can generalize the results  Eq. \ref{gen/UE} , Eq. \ref{gen/logit} as

\begin{equation}\label{generalized}
    u(t) = A(\xi(t), \lambda(t)) \omega (t) + B(\xi(t), \lambda(t)).
\end{equation}
From Lemma \ref{lemma/A} we have $A\geq0$. 

\begin{lemma}\label{lemma/uxi}
 For the generalized lane choice model and for $\omega(t) \geq 0$, we have $\frac{\partial u}{\partial \xi} > 0$.
\end{lemma}

\begin{proof}
 From Eq. \ref{eq/Axi}  and Eq. \ref{eq/Dxi}  $\omega(t) \frac{\partial A}{\partial \xi} + \frac{\partial B}{\partial \xi} >0$ for $\omega(t) \geq 0$.
\end{proof}

Note that in this generalized lane choice model based on VBM, $A$ and $B$ depend on both state variables, i.e., the residual service rate and the excess density.

\begin{figure*} 
\centering
	\includegraphics[width=0.9\textwidth]{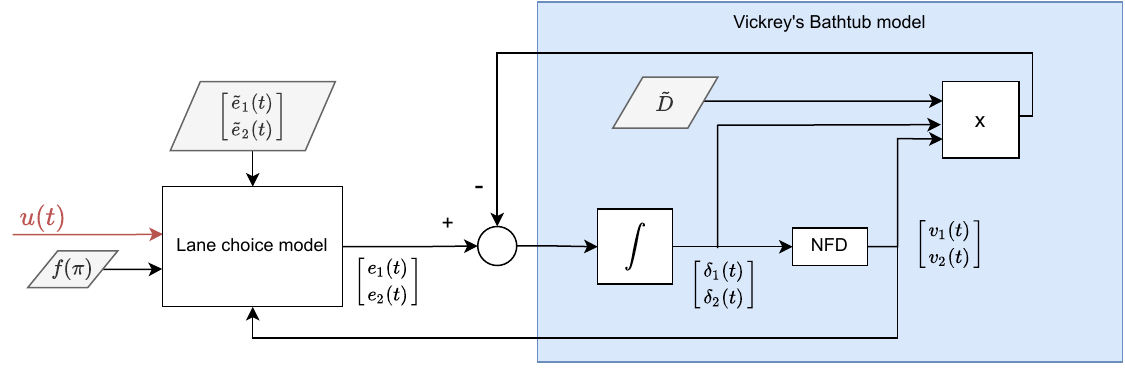} 
	\caption{\small Plant dynamics of the system where $u(t)$ is the distance-based  toll. The inputs to the plant are shaded in grey, they represent  the VOT distribution $f(\pi)$, the trip initiation rate of HOVs and SOVs, and the average trip distance $\tilde D$. } \label{fig:plant} 
\end{figure*}

In summary, the HOT lane system can be described with a traditional block diagram as in Figure \ref{fig:plant}. Note that the Figure is presented as an open-loop controller where the distance-based toll $u(t)$ is exogenous. In the next section, we will discuss the operational objectives of building a closed-loop system where $u(t)$ depends on the traffic state.



\section{Operation of HOT lanes}\label{sec/operation}

In this section, we present the two problems that operators face: (i) the definition of a distance-based congestion pricing scheme and (ii) the estimation of the lane choice model parameters. In reality, the operator does not know the drivers' VOT or their choice behavior. In the traditional HOT lanes pricing design framework, the system operators need to estimate the drivers' willingness to pay  \cite{YinandLou2009, Lou2011, Wang2020}, to propose a pricing scheme that encourages the desired number of SOVs to choose HOT lanes. However, we follow the approach proposed recently by Jin et al. \cite{Jin2020_HOT}, where the control problem is solved first, and the estimation problem is solved after that. This approach is different from the traditional way because the controller is independent of the lane choice model,  {built without assuming any particular choice model. Instead, the pricing strategy is traditionally based on an assumed lane choice model, e.g., multinomial logit}  \cite{YinandLou2009, Lou2011, Wang2020}.

In the following, we relate the two-fold operational objective (A2) to the bathtub model traffic dynamics.
The first operational objective interpretation, i.e., ``no congestion in the HOT lanes'', depends on the traffic model assumed. 
For the bathtub model, the first operational objective corresponds to the SUC or C states, based on Definition \ref{def/soc},  i.e.,
\begin{subequations}\label{eq/operationobj}
\begin{equation}
    \rho_1(t) \leq \rho_c l_1.
\end{equation}

The second operational objective in assumption A2, i.e., ``HOT lanes should be fully utilized'',  can be interpreted as maximizing the vehicles miles traveled in the HOT lanes during a certain time period $[0, T]$, and can be mathematically written as
\begin{equation}
    \max  \int_0^T L_1 q_1(t) dt,
\end{equation}
s.t.
\begin{equation}
    q_1(t) \leq C_1,
\end{equation}
\end{subequations}
where $q_1(t)$ is the internal flow rate measured in the HOT lanes, $L_1$ is the lane miles of the HOT lane, and $C_1$ is the capacity of the HOT lanes. We can achieve maximum efficiency of the HOT lanes when $q_1(t)=C_1$ at all times during the time $[0 , T]$. This operational objective is generic and independent of the traffic dynamics assumed.

For the VBM, the operational objectives presented (Eq. \ref{eq/operationobj}) are equivalent to operating the HOT lanes bathtub at the critical state, i.e., $\rho_1(t) = \rho_c l_1$ and $\lambda(t) =0$. This strategy is similar to the operational objectives for ramp metering control or variable speed limit control  \cite{Alinea, HEGYI2005_PartC, Martinez2020}, where the operational point with the maximum flow is targeted. 
Further,  we can impose $\dot \lambda(t) =0$  to ensure that the system stays at the desired operating point, i.e., impose a  time-independent active number of trips in the HOT lanes. From Eq. \ref{def:xi_lan}, this additional objective is equivalent to $\xi(t)=0$. 

Naturally, the operators can only achieve these objectives by modifying the toll for HOT lanes and indirectly increasing or decreasing the proportion of paying SOVs. From a control perspective, the reference signal is 
\begin{equation}\label{eq/reference}
    r(t) = \begin{bmatrix}
\lambda_0 \\
\xi_0
\end{bmatrix} = \begin{bmatrix}
0\\
0
\end{bmatrix} .
\end{equation}

\begin{figure*}
\centering
	\includegraphics[width=0.75\textwidth]{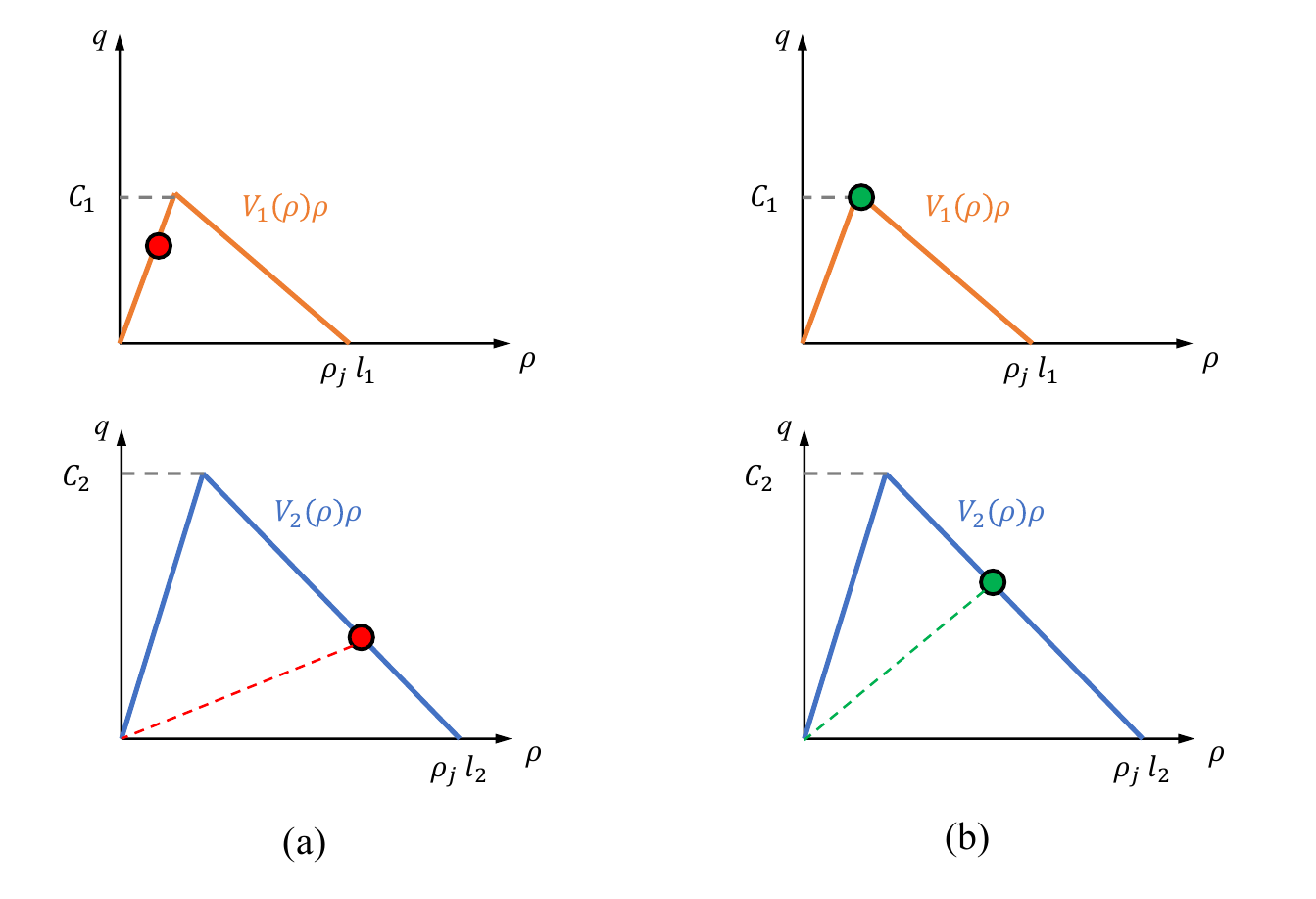} 
	\caption{\small The NFD for each type of lanes (upper diagram for HOT lanes and lower diagram for GP lanes). (a) Red dots represent operational points in each bathtub at a non-optimal state (b)  Green dots represent operational points in each bathtub under optimal congestion pricing.} \label{fig:operation} 
\end{figure*}

In the following, we discuss the relation between the two-fold operational objective (A2), which corresponds to Eq. \ref{eq/reference} for the bathtub model, to a maximization of the completion rate of trips across all lanes. For any given total number of vehicles on the corridor, we can define their distribution across lanes to achieve the maximum outflow. Intuitively, the maximum outflow on the corridor is achieved when no lane is under-utilized.

\begin{lemma}\label{lemma/maxflow}
 For the VBM with a triangular fundamental diagram, the maximum outflow given assumption A1 is achieved in three cases: Case I when $\rho_1=\rho_c$ ; Case II, when $\rho_2=\rho_c $, or Case III when all lanes are strictly congested. 
\end{lemma}


\begin{corollary}\label{corollary/maxflow}
 \textcolor{black}{The only case where VBM with a triangular fundamental diagram leads to the maximum outflow given assumption A1 that is consistent with the "full utilization of HOT lane(s)" (i.e., assumption A2) is Case I. }
\end{corollary}

The proof of Lemma \ref{lemma/maxflow} is presented in Appendix \ref{app:lemmamaxflow}.
\textcolor{black}{From Lemma \ref{lemma/maxflow} there are three cases that can lead to maximum flow. Note that Case III corresponds to a $u(t)=0$, which is equivalent to have no managed lane. Further, Corollary \ref{corollary/maxflow},   highlights that the } $\lambda(t) =0$ objective corresponds to maximization of trip completion rate in the HOT lanes for VBM. 
Graphically, it is clear that eliminating the excess density in the HOT lanes leads to maximum flow rate, as can be observed in Figure \ref{fig:operation}(b). \textcolor{black}{The results can be} compared to the fundamental diagrams under \textcolor{black}{a HOV operation of the managed lane, where the HOV lane(s) are underutilized,} Figure \ref{fig:operation}(a).  \textcolor{black}{The benefits of implementing a HOT operation for HOV lanes that are underutilized will be discussed through numerical results in Section \ref{sec:numerical}.}

\subsection{Congestion pricing scheme}\label{sec/control}

\begin{figure*}[t]
\centering
	\includegraphics[width=\textwidth]{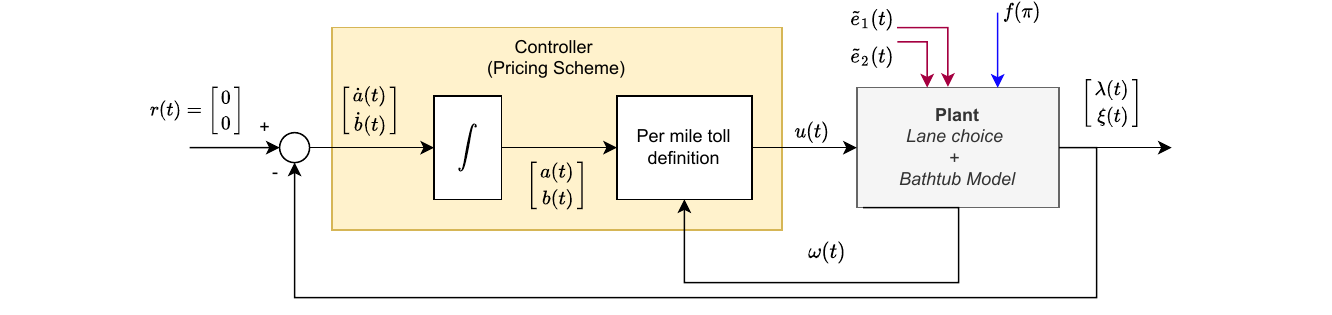}
	\caption{\small Block diagram of the control system. The distance-based toll $u(t)$ is defined by Eq. \ref{eq:u}, where $\omega(t)$ is per unit distance travel time difference between lanes and $a(t)$ and $b(t)$ are obtained from the integral and proportional controllers, respectively.} \label{fig:controller} 
\end{figure*}

The operator can not control the number of vehicles in the HOT lanes in a direct way. Therefore (s)he needs to define the pricing scheme $u(t)$ to control the system by modifying the lane choice behavior of SOVs. 
From a control perspective, the HOT lane pricing can be understood as a system that is controlled by an operator, i.e., charging a distance-based toll $u(t)$. 
The control system is presented in Figure \ref{fig:controller}, and the plant dynamics include the lane choice model and the bathtub model, as depicted in Figure \ref{fig:plant}. 
The pricing scheme proposed in this paper is based on a feedback control method, where the residual service rate and excess density need to be measured in real-time and fed back to the controller to update the dynamic price. Notice that even if the operator aims to modify the lane choice of users, the lane choice model itself is not needed for the controller to work; only the speeds of the bathtub and the state variables are fed back into the controller. 

The proposed toll should be a distance-based pricing scheme, as per Lemma \ref{lemma/mileage}. Moreover, from Eq. \ref{generalized} the distance-based toll should increase with the travel time difference rate, $\omega(t)$. Therefore, the proposed toll is 
\begin{subequations}\label{eq:controlab}
\begin{equation}\label{eq:u}
    u(t) = a(t) \omega(t) + b(t) ,
\end{equation}
\noindent
where $a(t)$ represents an hourly price [\$/h] and $b(t)$ is a distance-based price [\$/miles]. These time-dependent coefficients are determined by integral controllers as
\begin{equation}
    \dot a(t) =  K_1 \lambda(t) - K_2 \xi(t),
\end{equation}
\begin{equation}
     \dot b(t) =  K_3 \lambda(t) - K_4 \xi(t),
\end{equation}
\end{subequations}
where $\lambda(t)$ is the excess density defined in Eq. \ref{def/lan}, and $\xi(t)$ is the residual service rate as in  Eq. \ref{def/xi}. Thus, the units of $K_1$  and $K_4$ are $\$ miles / veh / h^2$, and $ \$/veh/miles$, respectively, and $K_2$ and $K_3$ are in $ \$ / h / veh$.
 {We will refer to this type of controller as a ``linear combination of I-controllers''. This controller is not common in the literature, and it has been specifically developed to serve this transportation system.}

The proposed controller is well behaved when all $K_i>0$, because when $\rho_1 >  \rho_{c}   $, i.e., HOT lanes are SOC, then the integral controllers Eq. \ref{eq:controlab} will lead to a increase of the toll $u(t)$. This will reduce the number of SOVs willing to pay if all other parameters stay constant. On the other hand, if $\rho_1 <  \rho_{c}   $, i.e., HOT lanes are SUC, the controller will lead to $\dot a (t),  \dot b (t) < 0$ reducing the toll to encourage more SOVs to use the HOT lanes. 
Further, the integral controller sign for $\xi(t)$ is chosen based on Lemma \ref{lemma/uxi}. If there is a residual service rate in the HOT lanes, $\xi(t) > 0$, the controller will reduce the price to reduce $\xi(t)$.

\subsection{Estimation of VOT}

The estimation of the parameters in the lane choice model (e.g., VOT distribution) is a relevant problem. The operator might be interested in that information even though the controller is independent of it. 
\textcolor{black}{Fortunately, given the relationships established in Section \ref{sec/lanechoice} between the distance-based pricing $u(t)$, the travel time differences per unit distance, and the proportion of SOVs choosing to pay,} the operator can estimate the VOT of users based on \textcolor{black}{observable variables}.
To do so, the operator needs to assume a lane choice model and measure the distance-based toll set, the difference in speeds between lane types, and the choices of the users\textcolor{black}{, which is consistent with previous work on a single bottleneck}  \cite{Jin2020_HOT}.

Assuming that the drivers choose their lanes based on cost minimization leading to the UE-based equilibrium (as in Section \ref{sec:UE}), one can estimate the VOT distribution from Eq. \ref{def/pUE}. At any given time $t$, based on the observed values $v_i(t)$, $\tilde e_2(t)$ and $\tilde e_{2,1}(t)$, and the operators distance-based cost, $u(t)$, the estimate of the cumulative distribution function follows
\begin{equation*}
    \hat F\left(\frac{u(t)}{\omega(t)} \right)= 1 - \frac{\tilde e_{2,1}(t)}{\tilde e_2(t)}.
\end{equation*}

Further, assuming that there is an error \textcolor{black}{ on the SOVs disutility} proportional to their trip distance that is unknown to the modeler (as in Section \ref{sec:logit}), one can estimate the VOT distribution from Eq. \ref{eq/pML}. At any given time $t$, based on the observed values $v_i(t)$, $\tilde e_2(t)$ and $\tilde e_{2,1}(t)$, and the toll, $u(t)$, the estimate of the cumulative distribution function follows
\begin{equation*}
    \hat {\pi^*}(t) = \frac{1}{\omega(t)} \left[ u(t) - \frac{\ln \left(\dfrac{\tilde e_2 }{\tilde e_{2,1}} -1 \right)}{\alpha_*} \right].
\end{equation*}
This estimation assumes that we have an estimate of $\alpha_*$, which is commonly considered to be 1.

Note that $\tilde e_1(t)$ (the demand of HOVs) is not needed to estimate the VOT, but the operator needs to measure the speeds in both types of lanes, $v_1(t)$ and $v_2(t)$, to calculate $\omega(t)$ from Eq. \ref{omega/def}. Note that theoretically, these speeds represent the speeds of the whole corridor. Thus, the speed should be measured at different points along the corridor to have a good estimate of $\omega(t)$. The errors from these estimated values can be caused by measuring errors, as well as delay errors, e.g., when the toll $u(t)$ is not updated frequently.

\section{Equilibrium state and stability of closed-loop system with proposed pricing scheme }\label{sec:EQHOT}


It is important to study the equilibrium point and stability of the closed-loop system analytically to ensure that the control proposed is well behaved, at least theoretically, under some idealized conditions. If it does not behave well analytically, on a real implementation, the systems could behave wildly without any practical benefits of implementing it.
In this section we consider a constant trip initiation rate of vehicles, i.e., $\tilde e_1(t)=\tilde e_1$ and $\tilde e_2(t)=\tilde e_2$, to study the equilibrium and stability of the proposed controller.

In Section \ref{sec:equilibrium} we describe the equilibrium solution of the closed loop system, assuming two different types of NFDs. In Section \ref{sec:stability} we prove the stability of the closed-loop system with the proposed controller.

\subsection{Equilibrium under constant demand}\label{sec:equilibrium}

The closed loop system is formed by the traffic dynamic ODEs of the VBM Eq. \ref{eq/dynamics/vbm}, the generalized lane choice model Eq. \ref{generalized} and the controller definition Eq. \ref{eq:controlab}. 

\begin{lemma}
The optimal state  $\lambda(t), \xi(t) = 0$ leads to equilibrium of the closed loop system.
\end{lemma}

\begin{proof}
From the optimal state $\lambda(t), \xi(t) = 0$ and Eq. \ref{eq:controlab} we have $\dot a (t), \dot b(t) = 0$. Further, from $\dot \xi(t) =0$ and Eq. \ref{def:xi_lan}  we have $\dot \lambda(t) =0$ and $\dot \delta_1(t) =0$.
\end{proof}

\begin{lemma}
The equilibrium of the closed loop system is the optimal state if $\dot a (t), \dot b(t) = 0$.
\end{lemma}

\begin{proof}
From the equilibrium assumption,  $\dot \lambda(t)=0$, $\dot \xi(t) =  0$, there is constant active number of vehicle-trips in the HOT lanes, i.e., $\dot \delta_1(t)=0$. From Eq. \ref{def:xi_lan} we have $\xi(t) = 0$, and from Eq. \ref{eq:controlab} we have  $\dot a(t)=K_1 \lambda_0$ and $\dot b(t)=K_3 \lambda_0$.  From $\dot a (t), \dot b(t) = 0$, clearly $\lambda(t)=\lambda_0=0$. Note that there might be an equilibrium point of the closed loop system with $\lambda(t)=\lambda_0\neq0$ and $\xi(t)=0$.
\end{proof}

In particular, at optimal equilibrium  we have $\delta_1 = L_1 \rho_c$ from Eq. \ref{def/lan}, and from Eq. \ref{nfd} the speed in the HOT lanes bathtub is $u_f$. Therefore, the traffic dynamic in the HOT lanes bathtub Eq. \ref{eq/dynamics/vbm} can be written as
\begin{subequations}
\begin{equation}\label{eq/p_VBM_1}
   0 = \tilde e_1+ \tilde e_2  p(t) - \frac{L_1}{D} \rho_c u_f .
\end{equation}
Thereafter,  the dynamics of the GP lanes bathtub can be written as
\begin{equation}\label{eq/ODE_del2_VBM}
    \dot \delta_2 (t) = \tilde e_1 + \tilde e_2 -  \frac{ L_1 \rho_c u_f +  \delta_2(t) V(\frac{\delta_2(t)}{L_2})}{D}.
\end{equation}
\end{subequations}
Notice that at the optimal equilibrium, there is a constant ratio of SOVs choosing to pay, i.e.,
\begin{equation}\label{eq/p_VBM}
   p_0 = \frac{L_1 \rho_c u_f -  \tilde e_1  D}{ D \tilde e_2 }.
\end{equation}
In the following, we discuss how the assumption of NFD impacts the equilibrium solution under a constant trip initiation rate.

\subsubsection{Triangular fundamental diagram assumption}

From assumption A1 we know that $\rho_2(t) > \rho_c$, and considering the speed-density relation in Eq. \ref{nfd}, the ODE Eq. \ref{eq/ODE_del2_VBM} leads to an exponential increase in density 
\begin{equation}\label{eq/exp_del2}
\begin{split}
    \delta_2(t) =& (\delta_2(0) + \frac{D \tilde e_2 (1-p_0)}{w} - \rho_j l_2 L) e^{wt/D} \\
    & -  \frac{D \tilde e_2 (1-p_0)}{w} + \rho_j l_2L,
\end{split}
\end{equation}
where $\delta_2(0)$ is the initial number of active trips in GP lanes at $t=0$.
The density in the GP lanes will increase until the jam density, where we will have $v_2(t)=0$. \textcolor{black}{Then, the cost of travelling in the GP lanes becomes infinite. Regardless of how much the price $u(t)$ increases, the cost of travelling in the HOT lane will always be finite, thus,} no more vehicles will be able to choose the GP, and everybody will choose the HOT lanes, eventually leading to gridlock too. In consequence, we conclude that assuming Eq. \ref{nfd} for constant trip initiation rate leads to gridlock independently of the controller. {This result is supported by domain knowledge since a steady state cannot materialize when the inflow to the corridor is higher than its capacity. }

In reality, gridlock is rarely observed because demand peaks over a short period. When demand decreases, the severe congestion may be recovered without reaching gridlock. {However,} the study of equilibrium and stability for time-dependent trip initiation rate, i.e., $\tilde e_1(t), \tilde e_2(t)$  is out of the scope of this paper. {In the future, if the departure-time choice can be modeled endogenously for the bathtub model, one could try to study the equilibrium of the HOT lane problem based on simulations.} 
Notice that the gridlock under constant trip initiation rate can be observed for other NFD assumptions that account for hypercongestion, i.e., decreasing flow with density.
 {Therefore, for this study, we need to rely on another assumption of the NFD to perform the equilibrium state and stability analysis.}


\subsubsection{Approximate triangular fundamental diagram assumption}

In the following, we study the equilibrium point for the closed-loop system under constant trip initiation rate for the assumption of ATFD, Eq. \ref{eq:ATFDspeed}. 
The density in the GP lanes will increase exponentially following Eq. \ref{eq/exp_del2} until a point where the flow is constant, $c$. We will refer to this time instant as $t_0$. After that, the flow is constant, 
and the ODE Eq. \ref{eq/ODE_del2_VBM} becomes
\begin{subequations}\label{rho2/equilibrium}
\begin{equation}
    \dot \delta_2(t) = c \omega_0, \text{ where,}
\end{equation}
\begin{equation}
    \omega_0 = \frac{\tilde e_2 (1- p_0)}{ c} - \frac{L_2}{D},
\end{equation}
\end{subequations}
leading to $\delta_2(t) = \omega_0 c t + \delta_2(t_0)$. From Eq. \ref{omega/def} and Eq. \ref{rho2/equilibrium} we have $\omega(t) = \omega_0 t + \frac{\delta_2(t_0)}{c} - \frac{1}{u_f}$. Thus, the per unit distance travel time difference also increases linearly in time and can be defined as
\begin{equation}\label{omega_equilibrium}
    \omega(t) =\omega_0 t + \omega_1 . 
\end{equation}
The distance-based toll Eq. \ref{eq:u} in equilibrium state also increases linearly in time as $ u(t) = a_0 \omega( t ) + b_0$. 






\subsection{Stability of equilibrium state}\label{sec:stability}

In the following we study the stability of the equilibrium point ($\lambda(t), \xi(t) = 0$) for the generalized lane choice model (Eq. \ref{generalized}) for VBM (Eq. \ref{eq/dynamics/vbm}). To do so, we will obtain a linear system and study the sign of its eigenvalues.

From  Eq. \ref{generalized} and Eq. \ref{eq:controlab}, we have 
\begin{equation*}
    a(t) + \frac{b(t)}{\omega(t)} = A(\xi(t), \lambda(t)) + \frac{B(\xi(t), \lambda(t))}{\omega(t)}.
\end{equation*}
Then, we take the temporal derivative on both sides, leading to
\begin{equation*}
\begin{split}
    &\dot a(t) + \frac{\dot b(t)}{\omega(t)}  - \frac{b(t) \dot \omega(t)}{\omega^2(t)}=\\ &\frac{\partial A}{\partial \xi} \dot \xi(t) + \frac{\partial A}{\partial \lambda} \dot \lambda(t) + \frac{1}{\omega(t)} \left( \frac{\partial B}{\partial \xi} \dot \xi(t) + \frac{\partial B}{\partial \lambda} \dot \lambda(t) \right)  - \frac{B(\xi,\lambda) \dot \omega(t)}{\omega^2(t)} .
\end{split}
\end{equation*}
\noindent From Eq. \ref{omega_equilibrium} we have $\dot \omega(t) = \omega_0$ and $\frac{\dot \omega(t)}{\omega^2(t)} \sim 0$ for large $t$. Thus, we can approximate the above expression by
\begin{equation*}
\begin{split}
    \dot a(t) + \frac{\dot b(t)}{\omega_0 t + \omega_1 }= & \left( \frac{\partial A}{\partial \xi} + \frac{1}{\omega_0 t + \omega_1 } \frac{\partial B}{\partial \xi} \right) \dot \xi(t)\\
    &+  \left( \frac{\partial A}{\partial \lambda}  + \frac{1}{\omega_0 t + \omega_1 } \frac{\partial B}{\partial \lambda} \right)\dot \lambda(t) .
\end{split}
\end{equation*}
Further, from Eq. \ref{def:xi_lan} and Eq. \ref{eq:controlab}  we have
\begin{equation}\label{aux}
\begin{split}
    &K_1 \lambda(t) - K_2 \xi(t)  + \frac{K_3 \lambda(t) - K_4 \xi(t) }{\omega_0 t + \omega_1 }= \\
    &\left( \frac{\partial A}{\partial \xi} + \frac{\frac{\partial B}{\partial \xi}}{\omega_0 t + \omega_1 }  \right) \dot \xi(t) - \left( \frac{\partial A}{\partial \lambda} + \frac{ \frac{\partial B}{\partial \lambda}}{\omega_0 t + \omega_1 } \right) \frac{\xi(t)}{L_1} .
\end{split}
\end{equation}
Again, for large $t$ the left hand side can be approximated by $K_1 \lambda(t) - K_2 \xi(t)$ close to the equilibrium. Then, defining $H(t) =  \frac{\partial A}{\partial \xi}  +  \frac{1}{\omega_0 t + \omega_1 }  \frac{\partial B}{\partial \xi}$, and $J(t) =\frac{\partial A}{\partial \lambda}  +   \frac{1}{\omega_0 t + \omega_1 } \frac{\partial B}{\partial \lambda} $, \ref{aux} can be re-written as
\begin{equation}\label{eq/xilinearized}
    \dot \xi (t) = \frac{1}{H(t)} \left(  K_1 \lambda(t) + ( \frac{J(t)- K_2 L_1}{L_1} - K_2 ) \xi(t) \right).
\end{equation}
From Lemma \ref{lemma/uxi} and Eq. \ref{omega_equilibrium}, we have that $H(t) > 0$.

From Eq. \ref{def:xi_lan} and Eq. \ref{eq/xilinearized} the linearized system is
\begin{equation}\label{eq/linearsystem}
\begin{bmatrix}
\dot \xi(t) \\ \dot \lambda (t)
\end{bmatrix}
=
\begin{bmatrix}
\frac{J(t) - K_2 L_1}{ L_1 H(t)}  & \frac{K_1}{ H(t)} \\
-\frac{1}{L_1} & 0
\end{bmatrix}
\begin{bmatrix}
 \xi(t) \\  \lambda (t)
\end{bmatrix} .
\end{equation}

Notice that the above feedback controller linearization is a non-standard  approach to nonlinear systems  \cite{Jin2020_HOT, Khalil2002}. The stability of the controller can be studied through the eigenvalues of the linear system.


\begin{lemma}\label{lemma/stability}
The linear system Eq. \ref{eq/linearsystem} is stable for SUC, for any $K_1, K_2 > 0$. Further, it is stable for SOC, if $K_2 > \frac{J(t)}{L_1}$.
\end{lemma}

The proof of Lemma \ref{lemma/stability} is presented in Appendix \ref{app:lemmaeigen}.

\begin{figure}
    \centering
    \includegraphics[width=0.3\textwidth]{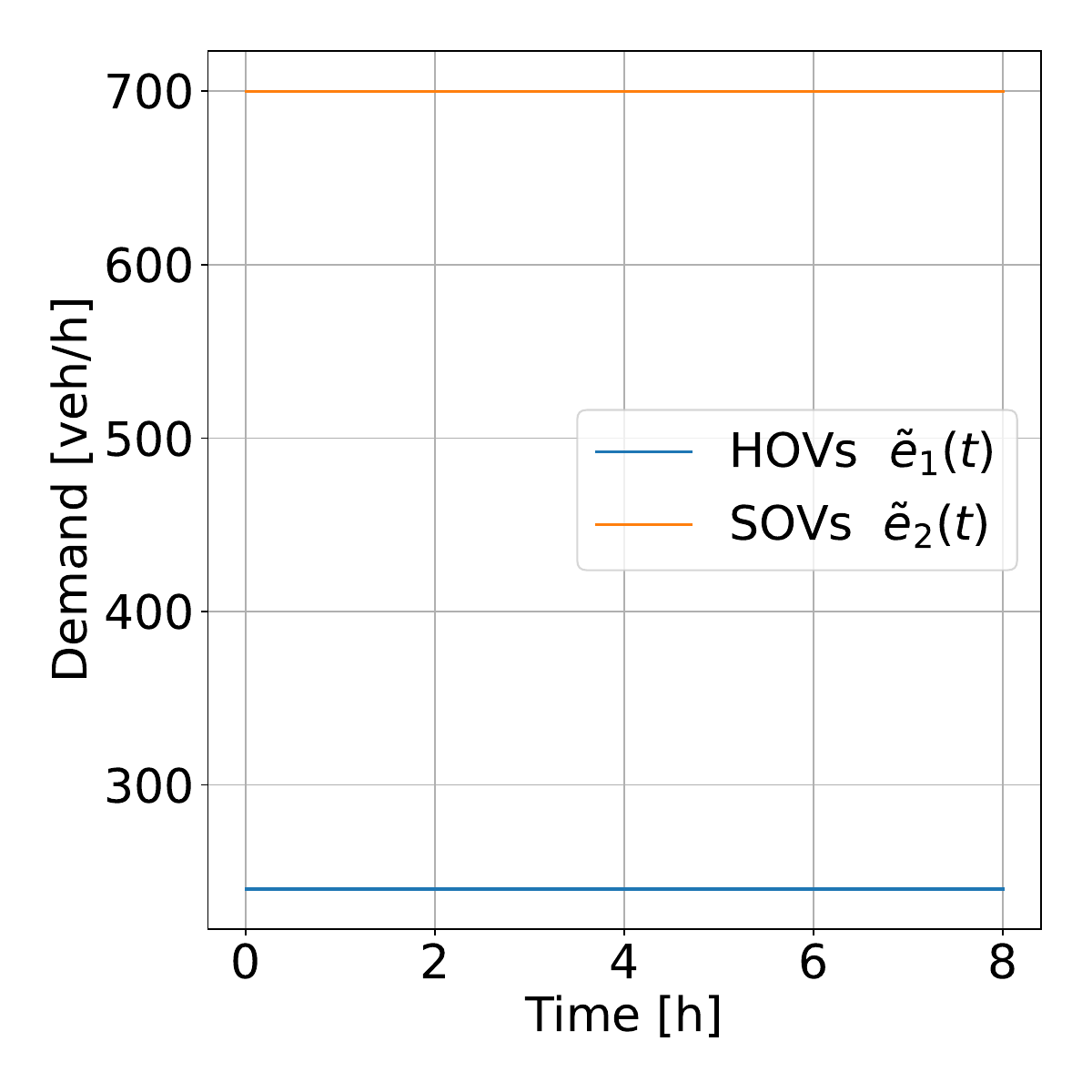}
            \caption{\small Demand pattern with constant trip initiation rate}\label{fig:demand-a}
\end{figure}

\begin{figure}
    \centering
    \includegraphics[width=0.3\textwidth]{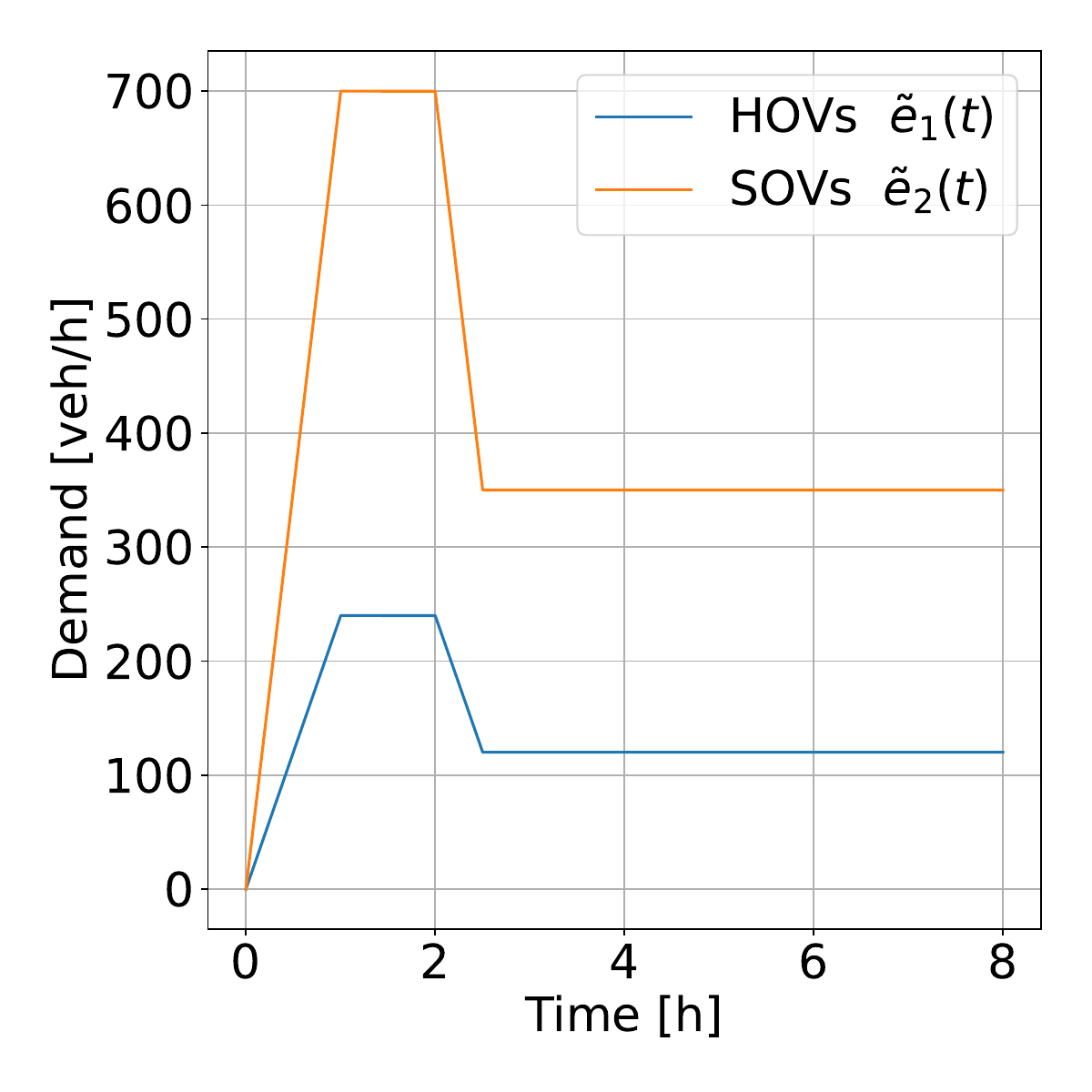}
    \caption{\small Demand pattern with short peak period.}\label{fig:demand-b}
\end{figure}

\section{Numerical examples}\label{sec:numerical}

In this section, we present several numerical examples of the closed-loop system with the proposed HOT lane pricing controller. In this section, we assume that the VOT distribution follows a negative exponential  distribution \cite{Jin2020_HOT}, following
\begin{equation*}
    f(\pi) = \frac{1}{E_{VOT}} e^{- \frac{\pi}{E_{VOT}}},
\end{equation*}
\noindent where $E_{VOT}=$ \$50/h is the expected VOT.

We will consider two trip initiation rate patterns, as shown in Figure \ref{fig:demand-a} and \ref{fig:demand-b}, where the mean trip distance is $\tilde D = 5$ km. Without loss of generality we assume that both bathtubs have one lane and the same parameters in the ATFD Eq. \ref{eq:ATFDspeed}, i.e., $u_f=100$ km/h, $w=20$ km/h, $\rho_j=$ 140 veh/km, and $c=0.8 C_0$.


\begin{figure*}
    \centering
    \includegraphics[width=\textwidth]{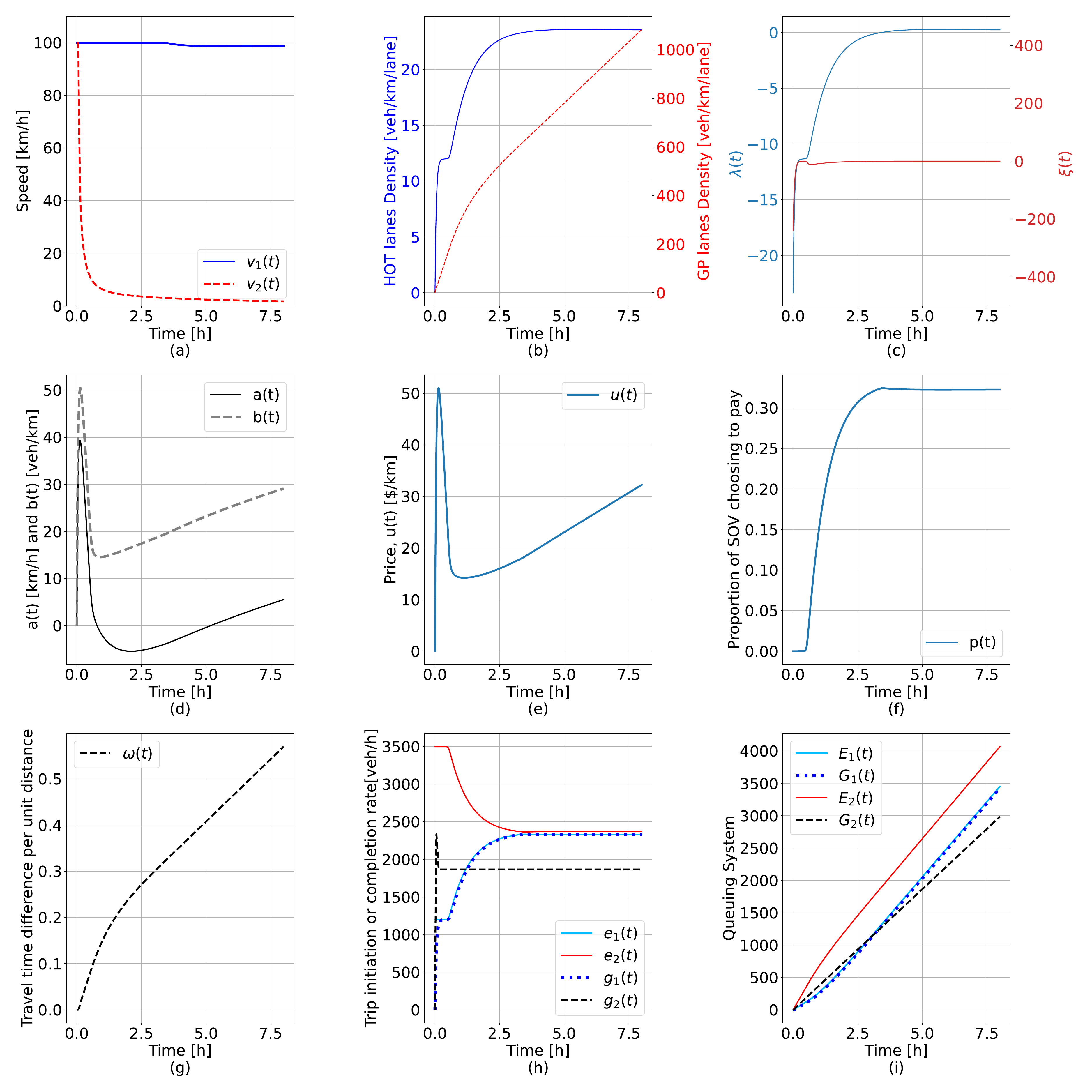}
    \caption{\small Simulation case under constant demand. Supply: ATFD with $c= 0.8 C_0$, lane choice model: UE with negative exponential distribution of VOT with expected value \$50/h. (a) Speeds, (b) Per lane-densities, (c) State variables, (d) $a(t)$ and $b(t)$, (e) Distance-based price, (f) Proportion of SOVs choosing to pay, (g) Per-mile travel time difference, (h) Inflow and outflows of each bathtub, (i) Queuing system view of each bathtub.}
    \label{fig:VBM_constant_demand}
\end{figure*}

\begin{figure*}
    \centering
    \includegraphics[width=\textwidth]{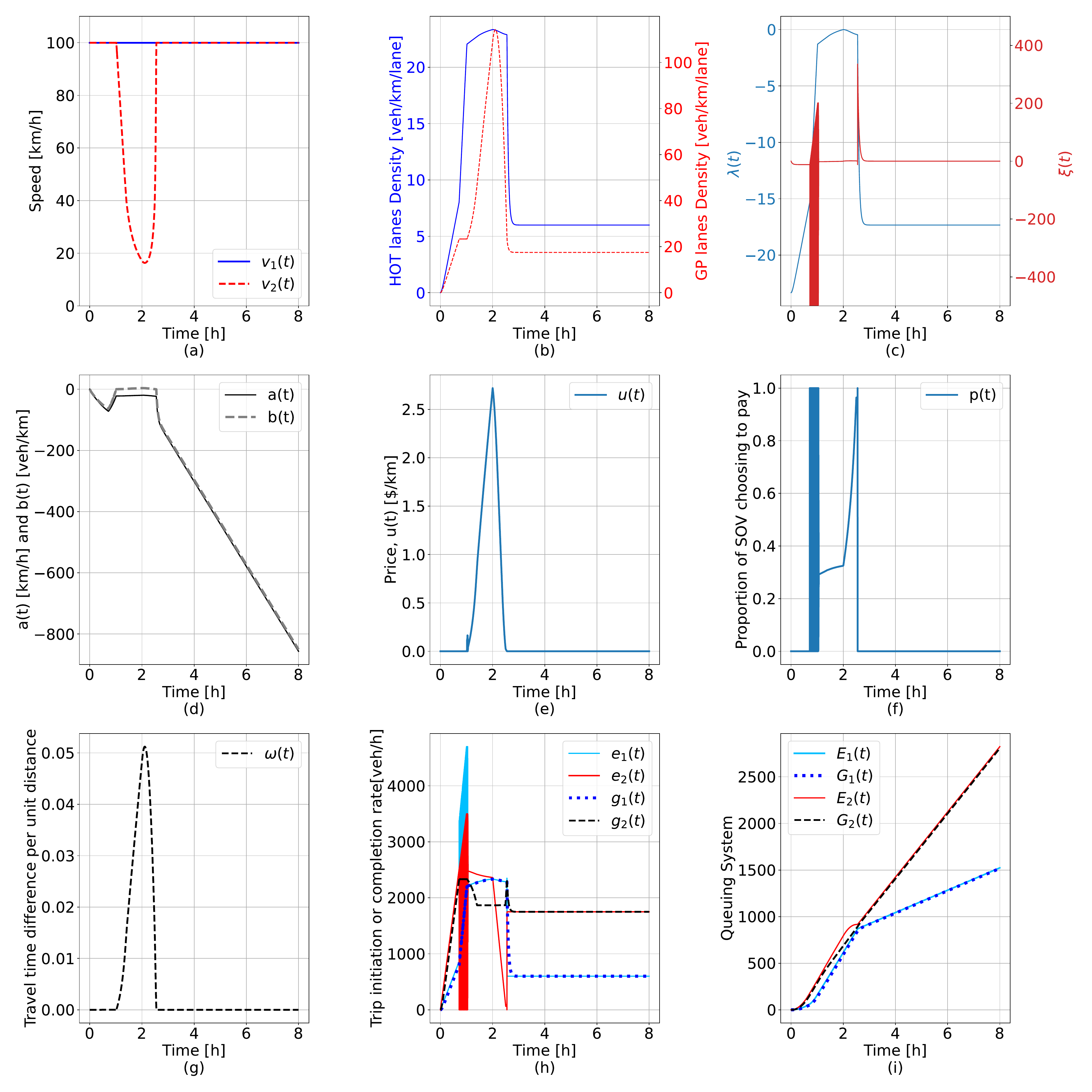}
    \caption{\small Simulation case under trapezoidal demand. Supply: ATFD with $c= 0.8 C_0$, lane choice model: UE with negative exponential distribution of VOT with expected value \$50/h. (a) Speeds, (b) Per lane-densities, (c) State variables, (d) $a(t)$ and $b(t)$, (e) Distance-based price, (f) Proportion of SOVs choosing to pay, (g) Per-mile travel time difference, (h) Inflow and outflows of each bathtub, (i) Queuing system view of each bathtub.}
    \label{fig:VBM_trapezoidal_demand}
\end{figure*}

The examples presented in this section correspond to a simulation with $\Delta t = 0.1 $s, and the controller in Eq. \ref{eq:controlab} has the following constants $K_1=K_3 =8$, $K_2= 5$, and $K_4=6$. \textcolor{black}{The parameters have been fine tuned with trial and error; for a systematic fine tuning of the $K_i$ the readers are referred to \cite{astrom1995pid}.}

First, we will assume a constant demand to validate the analytic results in Section \ref{sec:EQHOT} numerically.  
Figure \ref{fig:VBM_constant_demand} presents an example for the VBM with constant trip initiation rate. Clearly, after some time the equilibrium is reached and is the optimal with $\lambda(t), \xi(t) = 0$, and constant $p(t)$ slightly over 0.3.




Then, we consider a trapezoidal demand pattern Figure \ref{fig:demand-b} to show that the controller works well under a traditional peak period. 
The idea is to analyze a more realistic demand pattern than a constant trip initiation rate and study whether the controller works well under a time-dependent trip initiation rate. \textcolor{black}{Although the departure time pattern for the bathtub model has been recently studied \cite{ARNOTT2018,Li2019Analysis, BAO2021}, incorporating departure time on top of the lane choice model is out of the scope of this paper. } For this reason, we consider a typical peak period increase and decrease of demand with a trapezoidal profile.  
We assume that the TDD is the same at any given time and only the number of trips initiated changes over time. 
In this simulation scenario with short peak demand, the equilibrium is not reached, but  $u(t)$ increases to compensate for the peak demand and then decreases when the demand starts to decrease at $t=2$ h (see Figure \ref{fig:VBM_trapezoidal_demand}). For longer peak periods, e.g., 5 hours, the same equilibrium point can be reached as in Figure \ref{fig:VBM_constant_demand}, but the results are not presented here for brevity.

\begin{figure}
    \centering
            \includegraphics[width=0.3\textwidth]{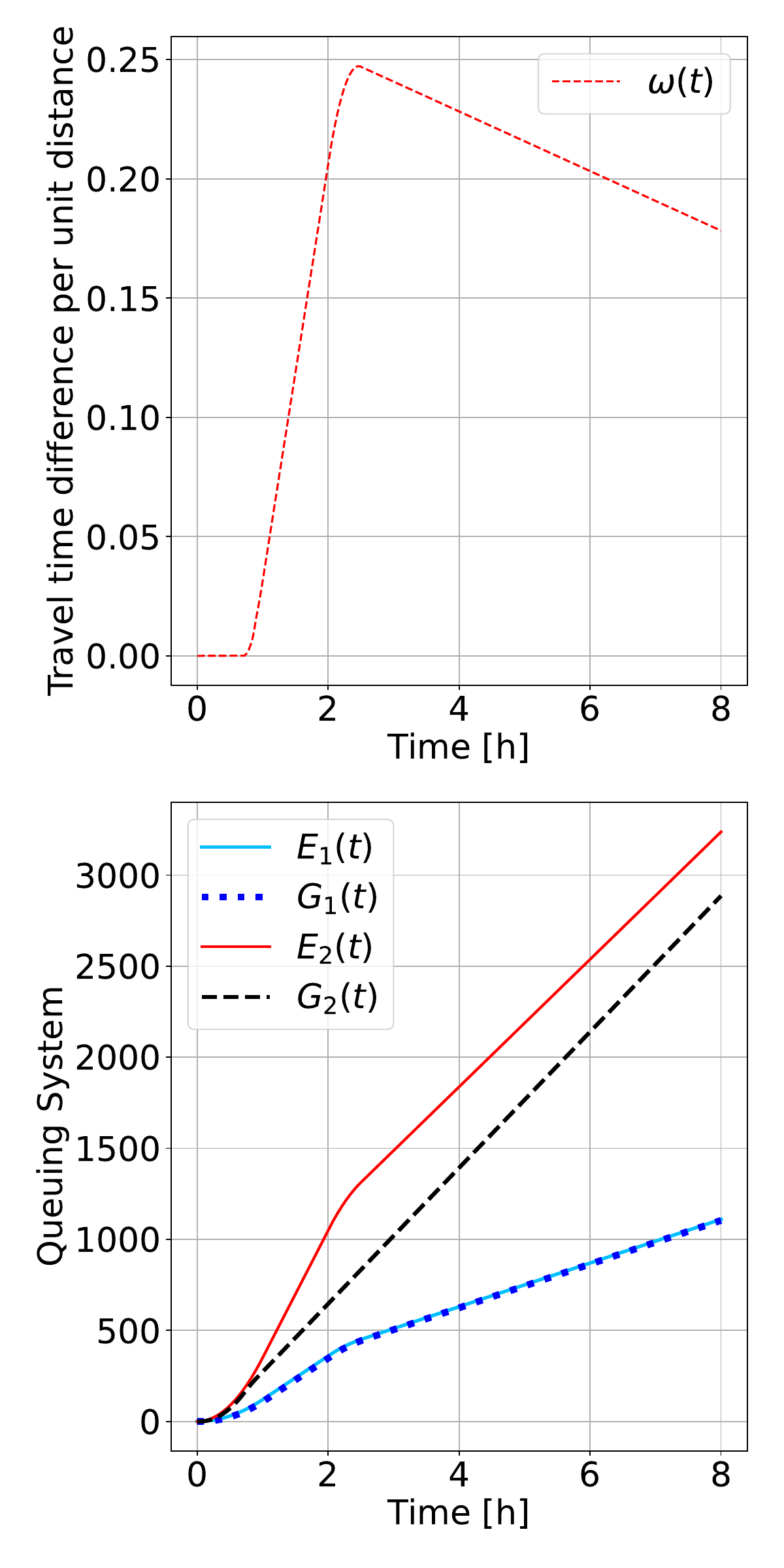}
            \caption{\small  HOV lanes scenario. Upper figures present the per-distance travel time difference $\omega(t)$, lower figures the queuing system for both the HOT lanes bathtub and the GP lanes bathtub.}\label{fig:comparison-a}
\end{figure}
\begin{figure}
    \centering
\includegraphics[width=0.3\textwidth]{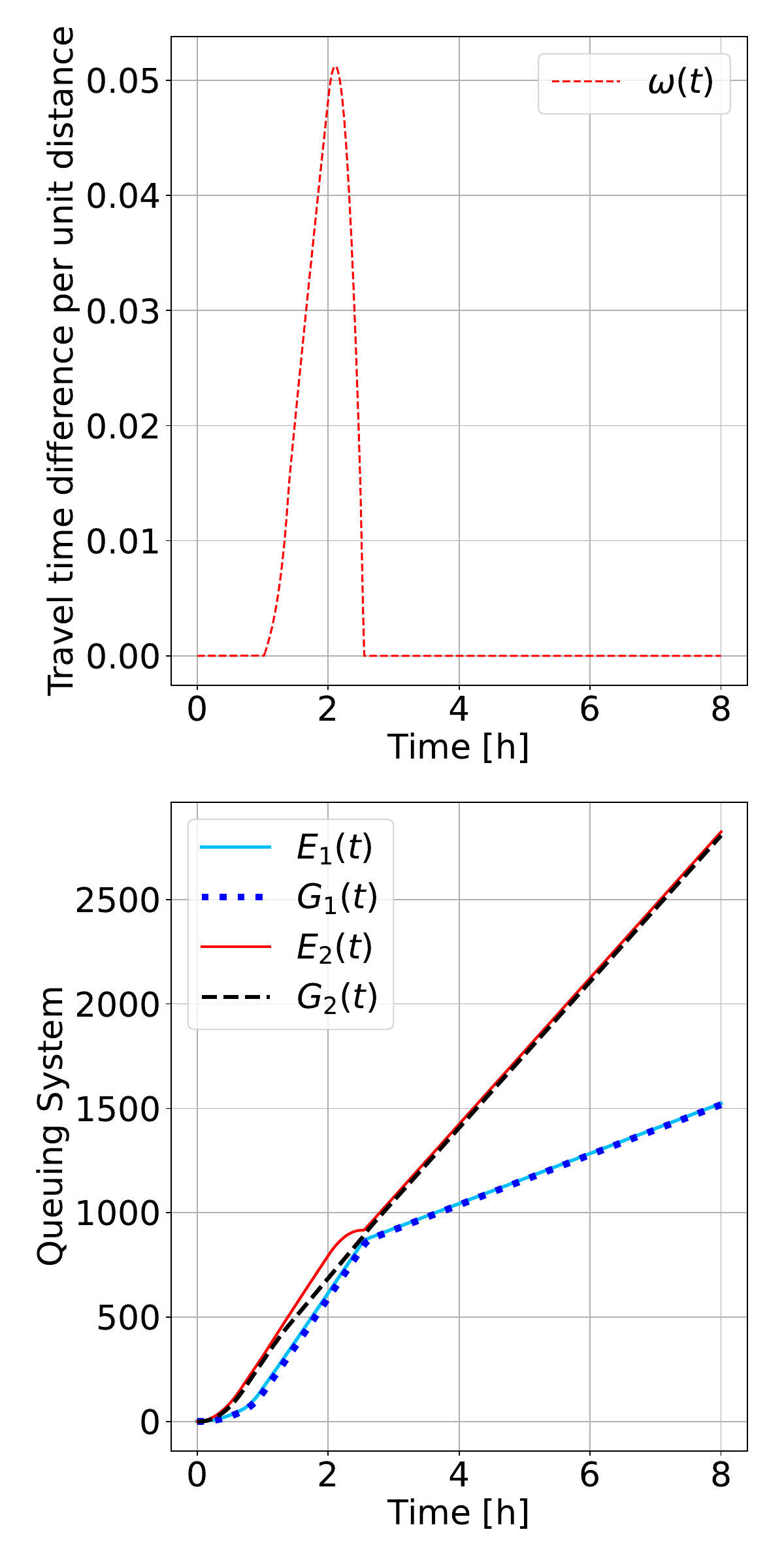}
    \caption{\small HOT lanes scenario with proposed controller. Upper figures present the per-distance travel time difference $\omega(t)$, lower figures the queuing system for both the HOT lanes bathtub and the GP lanes bathtub.}\label{fig:comparison-b}
\end{figure}

Besides verifying through numerical simulations that the proposed strategy is well-behaved, the numerical examples allow quantifying the benefits of the proposed HOT lane pricing strategy. To do so, we compare the proposed controller with the case where the HOT lanes work as HOV lanes, i.e., the price is effectively too large for any SOV to choose HOT lanes. We consider a short demand period with inflow for SOVs and HOVs as presented in Figure \ref{fig:demand-b} to compare the results. As shown in Figures \ref{fig:comparison-a} and \ref{fig:comparison-b}, the implementation of HOT lane pricing significantly reduces total delay in the system and the travel time difference among lane types. The flow rate in the GP lanes is improved, and the congestion period in the GP lanes is reduced to less than two hours. Note that since the average trip distance is $\tilde D = 5$ km, the vehicles in the GP will have, on average, a trip distance up to 1.2 hours longer than the trips in the HOV lanes under no HOT lane pricing control. In contrast, when HOT lane pricing is implemented, the largest mean trip travel time difference is 15 min. Furthermore, the HOT lanes are more utilized in the controlled case, and almost 50\% more vehicles can be served with the HOT lanes. 


\section{Conclusions}\label{sec/discussion}



In this paper, we proposed to use Vickrey's bathtub model (VBM) to capture traffic dynamics at a freeway corridor with multiple bottlenecks in a mathematically tractable way  \cite{Vickrey1991, Vickrey2020congestion}. Relying on this network trip flow model, we propose to study the high-occupancy-toll (HOT) lane pricing at the corridor level analytically.
We modeled the corridor system with two bathtubs: the HOT lanes for the HOVs and SOVs choosing to pay and the GP lanes for the SOVs that decide not to pay. We defined two key variables: the excess density in the HOT lanes, $\lambda(t)$, and the residual service rate, $\xi(t)$, allowing us to define the operational objective mathematically that satisfies assumption A2.
We assumed three different lane-choice models, including a deterministic vehicle-based UE model assuming that he VOT of SOVs’ follow a distribution, a stochastic UE assuming the logit model with fixed VOT for all SOVs, and a general lane-choice model. Further, we showed that distance-based tolling is sufficient to ensure a well-defined lane choice model through Lemma \ref{lemma/mileage}.  Then, we proposed a distance-based time-dependent toll as a linear combination of I-controllers. 
The analytical results in Section \ref{sec:EQHOT} show that the closed-loop system with constant demand has a stable and optimal equilibrium point for a generalized lane choice model and the VBM. In equilibrium, the proportion of SOVs choosing to pay is constant, and the density in the GP lanes and the distance-based toll increase linearly with time. 
Finally, the numerical examples presented in Section \ref{sec:numerical} verified the analytic results and showed that the controller also behaves adequately for time-dependent demand patterns, such as a typical peak demand period. The numerical examples also highlighted the compelling benefits of implementing HOT lane pricing on HOV lanes.

In summary, the contributions of this paper are three-fold. 
First, we formulated the HOT lane pricing problem for a freeway corridor with multiple origins and destinations with VBM  \cite{Vickrey1991, Vickrey2020congestion}. Second, we proposed a simple distance-based pricing scheme. Operators can implement this linear combination of I-controllers in real-world corridors by measuring only average speeds, the average density in the HOT lanes, and the total inflows and outflows to the HOT lanes.  Third, we proved that the proposed scheme has an optimal equilibrium point that satisfies the operational objectives in assumption A2 and is asymptotically stable.

In this paper, we proposed a time-dependent distance-based pricing scheme independent of the traffic flow model. Given the mathematical tractability of the VBM, we proved that such a controller is well-defined and stable under the assumption of VBM. 
In the future, we are interested in studying whether these desired characteristics of the proposed controller are transferable. A similar controller was developed by Jin et al. \cite{Jin2020_HOT} as a use-based toll for HOT lane pricing at a single bottleneck. Therefore, we expect the linear combination of I-controllers to be a suitable feedback control strategy given any underlying plant dynamics. It is possible that such a controller may be effective on other applications beyond HOT lane pricing. 
For these reasons, we are interested in considering the generalized bathtub model (GBM)  \cite{jin2020generalized}, where the trip distance distribution may not be a time-independent negative exponential, and other traditional traffic flow models, such as the CTM \cite{DaganzoCTM}, as the plant dynamics. We are also interested in making the trip initiation rate endogenous instead of assuming an exogenous demand to capture how the pricing strategy and the congestion level induce or suppress demand.

Further, this paper presents a framework to study HOT lane pricing from the corridor perspective that opens new research directions. For example, management strategies can be developed in the future to achieve a fairer system, where the distance-based cost $u(t)$ depends on the future states of the traffic system. To that end, the proposed model can be extended to account for the expected travel time of users, e.g., based on day-to-day information  {or near-future demand prediction. Then, the toll rate could account for} the overall expected externalities caused by each vehicle. For example, a vehicle starting the trip at the beginning of the peak period will have a higher marginal impact on traffic congestion than a vehicle starting the trip at the end of the peak period. For this reason, the toll for the first vehicle should be higher than for the second, even if the instantaneous per-mile travel time difference is the same.
This extension could be based on an agent-based approach of the bathtub model  \cite{Martinez2020Efficient}.

Another important research direction is to relax assumption A4, assuming that users do not stick to a single lane type for the entire trip. Instead, drivers may switch from GP to HOT lanes (or vice versa) several times during their trip. Under this more realistic assumption, multiple entrances and exits to the HOT lanes should be considered, similar to Pandey et al. \cite{Pandey2020}. For example, when the GP lanes are most congested, SOVs travel in HOT lanes and are willing to pay, but if the congestion mitigates their desired lane changes. In that case, SOVs would only pay for a part of their trip, corresponding to the proportional distance using the HOT lanes. Considering the GBM model with a discrete distribution of trip distances might be helpful to tackle the problem in a computationally more efficient way than relying on the CTM for the simulations. More importantly, with the GBM, the location to enter/exit the HOT lanes may be a continuous decision variable instead of considering the decision only at specific points.

\section*{Appendix: Proofs of Lemmas}

\subsection*{Proof of Lemma IV.1}\label{app:lemmamaxflow}

\begin{proof}
For any number of total vehicles in the road $\delta_1(t) + \delta_2(t)$, we define $\rho_{tot}(t) =\frac{\delta_1(t) + \delta_2(t)}{L}$. Thus, we can write the completion rate of trips as a function of $\rho_1(t)$, where $\rho_2(t) = \rho_{tot}(t) - \rho_1(t)$. Assuming $l_1 = l_2 = l$ without loss of generalization, from Eq. \ref{nfd}, the outflow can be written as
\begin{equation*}
\begin{split}
        g(t) = \frac{L_0 l}{D} \bigg(  & \min \left\{ \rho_1(t) u_f , w \left( \rho_j - \rho_1(t) \right) \right\} +\\
        & \min \left\{ u_f (\rho_{tot}(t) - \rho_1(t)) , w \left( \rho_j - \rho_{tot}(t) + \rho_1(t) \right) \right\} \bigg).
\end{split}
\end{equation*}

When, the HOT lanes are uncongested, $g_A(t) = \frac{L_0 l}{D} \bigg(  \rho_1(t) (u_f + w) +  w \left( \rho_j - \rho_{tot}(t) \right)  \bigg)$, which clearly increases in $\rho_1$.

When, the GP lanes are uncongested, $g_B(t) = \frac{L_0 l}{D} \bigg(  w \rho_j +  u_f \rho_{tot}(t) - \rho_1(t) (u_f  + w )   \bigg)$, which clearly decreases in $\rho_1(t)$.

When, all the lanes are congested, $g_C(t) = \frac{L_0 l}{D} \bigg( 2 w \rho_j  - \rho_{tot}(t) w  \bigg)$, is independent of $\rho_1(t)$.

When the HOT lanes are at the critical density, i.e., $\rho_1(t)=\rho_c $, the outflow $g_A(t)=g_C(t)$. Further, when the GP are at the critical density $g_B(t)=g_C(t)$. This can be easily verified with the critical density definition of $\rho_c=\frac{\rho_j w}{u_f + w}$, and is omitted here for brevity.

Thus, for VBM the maximum outflow given assumption A1 is $g_C(t) $ and can be achieved with $\rho_1=\rho_c $,  $\rho_2=\rho_c $ or with all lanes being congested.
\end{proof}

\subsection*{Proof of Lemma V.3}\label{app:lemmaeigen}

\begin{proof}
 The eigenvalues of \refe{eq/linearsystem}  are:

\begin{equation*}
    \frac{J(t) - K_2 L_1}{2 H(t) L_1} + \frac{1}{2} \sqrt{ \left( \frac{J(t) - K_2 L_1}{ H(t)L_1} \right) ^2 - 4 \frac{K_1}{L_1 H(t)}}
\end{equation*}
\begin{equation*}
    \frac{J(t) - K_2 L_1}{2 H(t)L_1} - \frac{1}{2} \sqrt{ \left( \frac{J(t) - K_2 L_1}{ H(t)L_1} \right) ^2 - 4 \frac{K_1}{L_1 H(t)}}
\end{equation*}

If $\left( \frac{J(t) - K_2 L_1}{ H(t)L_1} \right) ^2 \geq 4 \frac{K_1}{L_1H(t)}$, the eigenvalues are real. The system is Lypanvov' stable if both eigenvalues are negative. Thus, the condition for stability is that $\frac{J(t) - K_2 L_1}{ H(t) L_1} < 0$. For SUC $J(t)$ is negative and the condition is satisfied for any $K_2 > 0$. For SOC, $J(t)$ is positive and the condition is only satisfied for $K_2 > \frac{J(t)}{L_1}$.

If $\left( \frac{J(t) - K_2 L_1}{ H(t)L_1}  \right) ^2 < 4 \frac{K_1}{L_1 H(t)}$, both eigenvalues are imaginary. The real part of the eigenvalues must be negative for the system to be asymptotically stable. Then, the condition is the same than above, i.e., $\frac{J(t) - K_2 L_1}{ H(t) L_1} < 0$.
\end{proof}


\section*{Acknowledgements}
We would like to thank the ITS-Irvine Mobility Research Program (SB1) for partial financial support, under the project entitled "Dynamic pricing for high-occupancy toll lanes along a freeway corridor" in 2019-20. This research has been partially funded by Henry Samueli Endowed Fellowship and the Graduate Dean’s Dissertation Fellowship. The first author would like to thank Prof. Xuting Wang and Prof. Faryar Jabbari for the useful discussions. The results and views are the
authors’ alone.

\pdfbookmark[1]{References}{references}
\bibliographystyle{apalike}

\bibliography{refhot}

\end{document}